\documentclass[onecolumn,12pt,draft]{ieeetran}
\usepackage{amsmath}
\usepackage[final]{graphicx}
\usepackage{amsmath,epsfig,cite}
\usepackage{graphicx}
\usepackage{amssymb}
\usepackage{mathrsfs}
\newtheorem{lem}{Lemma}
\newtheorem{thry}{Theorem}

\begin{document}
\title{Asymptotic Capacity Analysis for Adaptive Transmission Schemes under General Fading Distributions}
\author{Yuan Zhang, Cihan Tepedelenlio\u{g}lu, \emph{Member, IEEE}
\thanks{The authors are with the School of Electrical, Computer,
and Energy Engineering, Arizona State University, Tempe, AZ 85287,
USA. (Email: yzhang93@asu.edu, cihan@asu.edu).} }
\date{}

\maketitle

\begin{abstract}
Asymptotic comparisons of ergodic channel capacity at high and low
signal-to-noise ratios (SNRs) are provided for several adaptive
transmission schemes over fading channels with general
distributions, including optimal power and rate adaptation, rate
adaptation only, channel inversion and its variants. Analysis of the
high-SNR pre-log constants of the ergodic capacity reveals the
existence of constant capacity difference gaps among the schemes
with a pre-log constant of $1$. Closed-form expressions for these
high-SNR capacity difference gaps are derived, which are
proportional to the SNR loss between these schemes in dB scale. The
largest one of these gaps is found to be between the optimal power
and rate adaptation scheme and the channel inversion scheme. Based
on these expressions it is shown that the presence of space
diversity or multi-user diversity makes channel inversion
arbitrarily close to achieving optimal capacity at high SNR with
sufficiently large number of antennas or users. A low-SNR analysis
also reveals that the presence of fading provably always improves
capacity at sufficiently low SNR, compared to the additive white
Gaussian noise (AWGN) case. Numerical results are shown to
corroborate our analytical results.
\end{abstract}

\begin{keywords}
Capacity, Power Adaptation, Fading Channels, Diversity
\end{keywords}


\section{Introduction} \label{introduction}

In order to perform spectrally efficient communications over fading
channels, adaptive transmission schemes are usually employed, which
use variable transmission power or rate according to the
instantaneous channel condition, possibly along with channel coding.
Determining the capacity of an adaptive transmission scheme under a
given fading channel distribution is of fundamental importance.
Specifically, ergodic capacity is usually used as a performance
benchmark for schemes over stationary ergodic channels. In
\cite{goldsmith97} the authors derive the optimal capacity with
channel state information (CSI) available at both the transmitter
and the receiver, which yields water-filling in time, and compare it
with two sub-optimal channel inversion (CI) and truncated channel
inversion (TCI) schemes. In \cite{laourine08} and \cite{laourine09}
ergodic capacity is derived for fading channels modeled with
generalized-k distribution, and G-distribution respectively. Both
fading models are parametric and describe practical composite
multi-path fading together with shadowing. Reference \cite{song05}
addresses ergodic capacity with multi-user diversity, while
\cite{alouini97,alouini99} analyze capacity under different
diversity combining schemes. Capacity under diversity combining in
the presence of spatial correlation and channel estimation error are
analyzed in \cite{shao99,bithas09,mallik04,khatalin07} and
\cite{hussain09} respectively. For implementation issues,
\cite{goldsmith97b} presents a class of adaptive uncoded $M$-ary
quadrature amplitude modulation (MQAM) schemes, and
\cite{vishwanath03} presents a class of adaptive turbo-coded
modulation schemes.

To the best of our knowledge, analytical comparisons of ergodic
channel capacity among adaptive transmission schemes for {\it
general} non-parametric channel distributions at high and low SNRs
have not yet been explored. In this paper, we provide asymptotic
capacity analysis at high and low SNRs for several schemes. We
consider optimal, and low-complexity sub-optimal adaptive
transmission schemes including CI and its variants, and compare them
with the corresponding AWGN capacity. By analyzing the high-SNR
pre-log constants (i.e. the limiting ratio of channel capacity to
the logarithm of the average SNR) of all the capacity curves, we
discover the existence of constant capacity difference gaps among
different schemes, and express these in closed form. Based on these
results, it is shown that fading always results in worse capacity
than AWGN at sufficiently high SNR regardless of the fading
distribution and the scheme used. Perhaps more surprising is that
arbitrary fading always improves ergodic capacity of the optimal
scheme at sufficiently low SNR, compared to the equivalent AWGN
channel. In addition, we confirm that the largest capacity
difference gap among all the schemes is the one between the CI
scheme and the optimal power and rate adaptation scheme.

Having expressions of gaps at high SNR between optimal and
sub-optimal schemes for general channel distributions, we consider
examples of when these sub-optimality gaps are reduced in the
presence of diversity. Clearly, the presence of diversity improves
channel capacity. What is less clear is whether the presence of
diversity closes the capacity difference gap at high SNR between
optimal and sub-optimal schemes. We provide the example of space
diversity in which it is proved that CI comes arbitrarily close to
the optimal capacity at high SNR, with sufficiently large number of
transmit/receive antennas. Another example proves the same with
sufficiently large number of users for the case of multi-user
diversity. The rate with which the sub-optimality gap reduces with
the number of antennas or users is also quantified, and it is shown
that antennas are more efficient than users in reducing this gap. We
also include comparisons among selected schemes which apply to the
low SNR region, showing analytically that AWGN capacity is always
exceeded by the optimal power and rate adaptation scheme under
fading at sufficiently low SNR, which is opposite to the high SNR
case already mentioned. Another interesting result establishes that
the presence of outage deteriorates the high-SNR capacity,
exhibiting a pre-log constant less than $1$. Ultimately these
results help to identify the trade-off between capacity and
complexity, since CI and TCI are known to require less complex
coding schemes than optimal power and rate adaptation
\cite{goldsmith97} to achieve the ergodic capacity. Our main
novelties, therefore, are in the consideration of general fading
distributions at high and low SNR regimes, and in identifying the
presence of diversity as a mechanism by which low-complexity CI
schemes are near-optimal.

The rest of the paper is organized as follows. Section
\ref{chnl_sys} establishes the channel and system model. Section
\ref{cap_anls} reviews the capacities of different adaptive
transmission schemes. Section \ref{asympt} provides asymptotic
analysis of these schemes, for both high and low SNR regions.
Section \ref{gap_div} analyzes how the high-SNR capacity difference
gaps behave under certain situations of diversity, and reveals
possible near-optimality of CI in the presence of diversity.
Simulation results are shown in Section \ref{simres}, and Section
\ref{concl} concludes the paper.

\section{Channel and System Model} \label{chnl_sys}

Consider the following model:
\begin{equation} \label{trans_rec}
y= \sqrt{\rho} s+ \nu
\end{equation}
where $\rho$ is the instantaneous channel power gain satisfying
E$[\rho]= \overline{\rho}< \infty$, $s$ is the transmitted sample,
$y$ is the received sample, and $\nu$ is AWGN with variance $N_0$.
Note that (\ref{trans_rec}) represents an equivalent fading channel
model capturing many situations other than single-input
single-output (SISO) systems with equivalent SISO characterizations,
such as multi-input single-output (MISO) and single-input
multi-output (SIMO) systems. Assume the average transmission power
is fixed to be E$[|s|^2]=S$. In this case, the received
instantaneous SNR becomes $\rho |s|^2/ \textrm{E} [|\nu|^2]= D
\gamma$, where $D \triangleq |s|^2/S$ denotes the instantaneous
power ratio, and $\gamma=S \rho/N_0$ denotes the instantaneous SNR
with constant power $S$. We assume $\gamma$ has probability density
function (PDF) $f_{\gamma}(x)$ and cumulative distribution function
(CDF) $F_{\gamma}(x)$. Notice that for adaptive transmission
schemes, $D$ may become a function of $\gamma$ since the
instantaneous power can be adapted to the channel. The ergodic
channel capacity is given by
\begin{equation}
C=\int_0^{\infty} \log (1+D(x) x) f_{\gamma}(x) d x= \textrm{E}
[\log (1+ D(\gamma) \gamma)]
\end{equation}
per unit bandwidth, where the logarithm is natural so that $C$ is in
nats per channel use. Note that $C$ depends on the average transmit
power $S$ through $f_{\gamma}(x)$.

The distribution of the instantaneous SNR random variable $\gamma$
is not assumed to be of any specific parametric form for our high
and low SNR results. We do however have some regularity conditions
and make one or both of the following assumptions on $\gamma$ for
the different results in the sequel:
\begin{itemize}
\begin{item}
{\bf A1}: $F_{\gamma}(x)$ is regularly varying at $0$
\cite[VIII.8]{fellerbook71}: $F_{\gamma}(x)=x^{d} \hspace{0.5mm}
l(x)$, where $0< d< \infty$ and $l(x)$ is a slowly varying function
at $0$, which by definition satisfies $\lim_{x \rightarrow 0} l(\tau
x)/l(x)=1$ for $\tau>0$;
\end{item}
\begin{item}
{\bf A2}: $0<\textrm{E}[\gamma^{-1}]< \infty$ (when {\bf A1} holds,
this occurs if $d>1$).
\end{item}
\end{itemize}
Assumption {\bf A1} implies that $F_{\gamma}(x)$ behaves like
$x^{d}$ near $x=0$, which can be shown to yield a diversity order of
$d$ (for a similar result please see \cite{wang03}). It also implies
that $F_{\gamma}(0)=0$ and $|\textrm{E}[\log \gamma]|< \infty$ which
are proved in Appendix \ref{reg_vary_proof}. Note that
$F_{\gamma}(0)=0$ requires that the fading distribution has no point
of mass at the origin which means that the PDF cannot have impulses
at the origin. {\bf A1} is satisfied by all channel distributions
considered in the literature and will be used in the sequel.
Assumption {\bf A2} ensures that the capacity of CI is nonzero and
holds in the presence of space or multi-user diversity. In the
sequel, {\bf A1} is necessary for most results, while {\bf A2} is
necessary for results pertaining to CI and TCI and not needed for
other results.

\section{Adaptive Transmission Schemes} \label{cap_anls}

In this section, we briefly review different adaptive transmission
schemes, since their expressions will be needed to derive our
asymptotic results. We begin with the capacity of the AWGN channel
with the same SNR as the average SNR of the fading channel in
(\ref{trans_rec}) which is E$[\gamma]=S \overline{\rho}/N_0$,
\begin{equation} \label{cap_awgn}
C_{\rm AWGN}= \log \left( 1+\frac{S \overline{\rho}}{N_0} \right).
\end{equation}
In what follows, we describe several schemes and formulate their
ergodic capacities, based on which we will perform asymptotic
analysis at high and low SNRs.

\subsection{Optimal Power and Rate Adaptation (OA)} \label{oa}

We first consider the optimal power and rate adaptation scheme
subject to normalized average power constraint $\textrm{E}
[D(\gamma)]=1$. As per \cite{goldsmith97}, the optimized $D(\gamma)$
is given by
\begin{equation}
D(\gamma)=\left( \frac{1}{\gamma_{\rm t}}- \frac{1}{\gamma} \right)
I[\gamma>\gamma_{\rm t}]
\end{equation}
where $\gamma_{\rm t}$ is a threshold determined by the average
power constraint
\begin{equation} \label{oa_pave}
\int_{\gamma_{\rm t}}^{\infty} \left( \frac{1}{\gamma_{\rm t}}-
\frac{1}{x} \right) f_{\gamma}(x) dx=1
\end{equation}
and $I[\cdot]$ is the indicator function. Clearly $\gamma_{\rm t}$
is uniquely determined, since the left hand side of (\ref{oa_pave})
is monotonically decreasing with $\gamma_{\rm t}$. The optimal
capacity becomes \cite{goldsmith97}
\begin{equation} \label{cap_oa}
C_{\rm OA}=\int_{\gamma_{\rm t}}^{\infty} \log \left(
\frac{x}{\gamma_{\rm t}} \right) f_{\gamma}(x) dx .
\end{equation}

\subsection{Rate Adaptation with Receive CSI only (RA)}

We now consider another scheme in which the instantaneous power is
independent of the channel (i.e. $D(\gamma)=1$ for any channel
realization). The capacity is simply given as
\begin{equation} \label{cap_cpra}
C_{\rm RA}=\int_0^{\infty} \log (1+x) f_{\gamma}(x) dx .
\end{equation}
As pointed out in \cite{goldsmith97}, (\ref{cap_cpra}) is applicable
as a benchmark capacity for schemes with receiver side CSI only,
provided that the input distribution which maximizes mutual
information is the same regardless of the channel state
\cite{mceliece84}. This holds for fading channels with AWGN
\cite{goldsmith97}. Note that $C_{\rm RA} \leq C_{\rm AWGN}$ due to
Jensen's inequality, while no such relation can be easily
established between $C_{\rm OA}$ and $C_{\rm AWGN}$ for all $S$.

\subsection{Channel Inversion (CI) with Variable Power and Constant Rate}

As its name indicates, CI is a scheme under which the transmission
power is adapted according to the channel so that the instantaneous
received SNR is kept constant. If {\bf A2} holds so that
E$[\gamma^{-1}] < \infty$, we define
$D(\gamma)=(\textrm{E}[\gamma^{-1}] \gamma)^{-1}$ so that
$\textrm{E} [D(\gamma)]=1$, which yields the constant instantaneous
received SNR, $\textrm{E}^{-1} [\gamma^{-1}]$, with the
corresponding capacity
\begin{equation} \label{cap_ci}
C_{\rm CI}=\log (1+\textrm{E}^{-1} [\gamma^{-1}]).
\end{equation}
Since CI effectively turns a fading channel into an AWGN channel,
any coding scheme that is suitable over AWGN channels would be
appropriate for CI, which is not the case for OA or RA. Therefore CI
is considered to be a viable sub-optimal scheme due to its low
complexity \cite{goldsmith97}.

\subsection{Truncated Channel Inversion (TCI)}

For the CI scheme we have described, {\bf A2} is needed for it to
have non-zero capacity. We now consider a variant of CI which is
applicable without {\bf A2} and has instantaneous power ratio given
by
\begin{equation} \label{otci_pinst}
D(\gamma)= \left\{ \begin{array}{ll}
0 & \textrm{if $\gamma<\gamma_{\rm t}$}\\
D_{\rm max} \gamma_{\rm t}/\gamma & \textrm{if $\gamma \ge
\gamma_{\rm t}$}
\end{array} \right. .
\end{equation}
In (\ref{otci_pinst}) the transmission is ceased when $\gamma$ is
below the threshold $\gamma_{\rm t}$, and $D_{\rm max}$ is related
to the threshold $\gamma_{\rm t}$ through the average power
constraint
\begin{equation} \label{otci_pave}
\int_{\gamma_{\rm t}}^{\infty} \frac{D_{\rm max} \gamma_{\rm t}}{x}
f_{\gamma}(x) dx=1.
\end{equation}
From (\ref{otci_pave}), we obtain $D_{\rm max}$ as a function of
$\gamma_{\rm t}$,
\begin{equation} \label{papr_otci}
D_{\rm max}(\gamma_{\rm t})= \left[ \int_{\gamma_{\rm t}}^{\infty}
\frac{\gamma_{\rm t}}{x} f_{\gamma}(x) dx \right] ^{-1}
\end{equation}
and the capacity is given by
\begin{equation} \label{cap_otci}
C_{\rm TCI}=[1-F_{\gamma}(\gamma_{\rm t})] \hspace{1mm} \log
(1+D_{\rm max} \gamma_{\rm t}).
\end{equation}

\subsection{Continuous-power Truncated Channel Inversion (CTCI)}

We now consider another variant of CI which, like TCI, is applicable
without {\bf A2}:
\begin{equation} \label{ctci_pinst}
D(\gamma)= \left\{ \begin{array}{ll}
D_{\rm max} & \textrm{if $\gamma<\gamma_{\rm t}$}\\
D_{\rm max} \gamma_{\rm t}/\gamma & \textrm{if $\gamma \ge
\gamma_{\rm t}$}
\end{array} \right. .
\end{equation}
Unlike (\ref{otci_pinst}), (\ref{ctci_pinst}) is a continuous
function of $\gamma$ and does not exhibit outage. In Section
\ref{asympt} the effect of outage on capacity at both high and low
SNRs will be clearly delineated. CTCI generalizes RA ($\gamma_{\rm
t}= \infty$) and CI ($\gamma_{\rm t}=0$), and will be useful in our
subsequent analysis to compare these schemes. Similar to TCI,
$D_{\rm max}$ is related to the threshold $\gamma_{\rm t}$ through
the average power constraint
\begin{equation} \label{ctci_pave}
D_{\rm max} F_{\gamma}(\gamma_{\rm t})+ \int_{\gamma_{\rm
t}}^{\infty} \frac{D_{\rm max} \gamma_{\rm t}}{x} f_{\gamma}(x)
dx=1.
\end{equation}
From (\ref{ctci_pave}), we obtain $D_{\rm max}$ as a function of
$\gamma_{\rm t}$ given by
\begin{equation} \label{papr_ctci}
D_{\rm max}(\gamma_{\rm t})= \left[ F_{\gamma}(\gamma_{\rm t})+
\int_{\gamma_{\rm t}}^{\infty} \frac{\gamma_{\rm t}}{x}
f_{\gamma}(x) dx \right] ^{-1}.
\end{equation}
Consequently, the capacity is given by
\begin{equation} \label{cap_ctci}
C_{\rm CTCI} = \int_0^{\gamma_{\rm t}} \log (1+D_{\rm max} x)
f_{\gamma}(x) dx+ [1-F_{\gamma}(\gamma_{\rm t})] \hspace{1mm} \log
(1+D_{\rm max} \gamma_{\rm t}).
\end{equation}

\section{Asymptotic Comparisons} \label{asympt}

We now compare the ergodic capacities of different schemes based on
their asymptotic properties at both high and low average SNRs. We
start by analyzing their high-SNR pre-log constants, and then
determine the constant capacity differences among those schemes that
have a high-SNR pre-log constant of $1$. We also provide comparisons
of selected schemes for low average SNR. In order to perform the
subsequent asymptotic analysis, we need to separate the
channel-dependent and channel-independent parameters which are
involved in the capacity formulae. Since $\gamma$ is a linear
function of $S$, we define the effective channel gain $z_{\rm
eff}:=\gamma/S$ whose PDF $f_{z_{\rm eff}}(z)$ and CDF $F_{z_{\rm
eff}}(z)$ are only related to the channel distribution, so that
$z_{\rm eff}$ does not depend on the average power $S$. Similarly,
we define the threshold $z_{\rm t}:=\gamma_{\rm t}/S$ to facilitate
our analysis.

\subsection{Asymptotic Slopes and Pre-log Constants of Capacity Curves} \label{asympt_slopes}

We now consider each of the schemes in Section \ref{cap_anls} to
determine how their capacities behave at both high and low SNRs. For
any capacity expression of the form $C=\textrm{E}[\log (1+S D(z_{\rm
eff}) z_{\rm eff})]$ we analyze the pre-log constant $\lim_{S
\rightarrow \infty} d C/d (\log S)$ as defined e.g. in
\cite{lapidoth05}. We also examine the low-SNR slope $\lim_{S
\rightarrow 0} d C/d S$ since in most cases $C$ becomes
approximately linear with $S$ as $S \rightarrow 0$. For the AWGN
capacity given by (\ref{cap_awgn}), we have the well-known
\begin{equation} \label{slope_awgn_sml}
\hspace{20mm} \lim_{S \rightarrow 0} \frac{dC_{\rm AWGN}}{dS}=
\frac{\overline{\rho}}{N_0}= \textrm{E}[z_{\rm eff}]
\end{equation}
\begin{equation}
\lim_{S \rightarrow \infty} \frac{d C_{\rm AWGN}} {d (\log S)}=1
\end{equation}
as a benchmark, where E$[z_{\rm eff}]$ is finite in
(\ref{slope_awgn_sml}) since we have assumed $\overline{\rho}<
\infty$. We proceed with the fading case starting with the optimal
power and rate adaptation.

\subsubsection{Optimal Power and Rate Adaptation}

For the case of OA, we rewrite (\ref{oa_pave}) and (\ref{cap_oa}) as
\begin{equation} \label{oa_pave2}
\int_{z_{\rm t}}^{\infty} \frac{1}{S} \left( \frac{1}{z_{\rm t}}-
\frac{1}{z} \right) f_{z_{\rm eff}}(z) dz=1
\end{equation}
\begin{equation} \label{cap2_oa}
C_{\rm OA}=\int_{z_{\rm t}}^{\infty} \log \left( \frac{z}{z_{\rm t}}
\right) f_{z_{\rm eff}}(z) dz .
\end{equation}
We express the slope and pre-log constant in terms of $z_{\rm t}$
since $C_{\rm OA}$ is implicitly related to $S$ through $z_{\rm t}$
in (\ref{oa_pave2}). Differentiating (\ref{cap2_oa}) with respect to
$z_{\rm t}$, we obtain
\begin{equation}
\frac{d C_{\rm OA}}{d z_{\rm t}}=\frac{F_{z_{\rm eff}}(z_{\rm t})-1}
{z_{\rm t}}.
\end{equation}
Solving for $S$ from (\ref{oa_pave2}) as a function of $z_{\rm t}$,
we obtain
\begin{equation} \label{sfn_zt}
Sz_{\rm t}=1-F_{z_{\rm eff}}(z_{\rm t})-z_{\rm t} \int_{z_{\rm
t}}^{\infty} \frac{1}{z} f_{z_{\rm eff}}(z) dz,
\end{equation}
and
\begin{equation} \label{diff_p_zt}
\frac{d z_{\rm t}}{d S}=\frac{z_{\rm t}^2}{F_{z_{\rm eff}}(z_{\rm
t})-1}.
\end{equation}
Consequently we have
\begin{equation} \label{oa_slope0}
\frac{d C_{\rm OA}}{d S}= \frac{d C_{\rm OA}}{d z_{\rm t}} \frac{d
z_{\rm t}}{d S}=z_{\rm t}
\end{equation}
and
\begin{equation} \label{oa_slope1}
\frac{d C_{\rm OA}}{d (\log S)}= S \frac{d C_{\rm OA}}{d S}= S
z_{\rm t}=1-F_{z_{\rm eff}}(z_{\rm t})-z_{\rm t} \int_{z_{\rm
t}}^{\infty} \frac{1}{z} f_{z_{\rm eff}}(z) dz.
\end{equation}
To determine the limits of (\ref{oa_slope0}) and (\ref{oa_slope1}),
we have the following lemma.
\begin{lem} \label{lem_zt}
The function $S \mapsto z_{\rm t}$ defined implicitly through
(\ref{oa_pave2}) has limits $\lim_{S \rightarrow 0} z_{\rm t}=\sup
\{ z: F_{z_{\rm eff}}(z) < 1 \}$ and $\lim_{S \rightarrow \infty}
z_{\rm t}=0$.
\end{lem}
\begin{proof}
See Appendix \ref{lem_zt_proof}.
\end{proof}
Note that for commonly considered fading channels, including
Nakagami-$m$, Ricean and log-normal, $z_{\rm eff}$ has an infinite
support and thus $\sup \{ z: F_{z_{\rm eff}}(z) < 1 \}= \infty$.
Using Lemma \ref{lem_zt} and (\ref{oa_slope1}) we have
\begin{equation}
\lim_{S \rightarrow \infty} \frac{d C_{\rm OA}}{d (\log S)}=
\lim_{z_{\rm t} \rightarrow 0^+} \left( 1-F_{z_{\rm eff}}(z_{\rm
t})-z_{\rm t} \int_{z_{\rm t}}^{\infty} \frac{1}{z} f_{z_{\rm
eff}}(z) dz \right).
\end{equation}
Clearly $\lim_{z_{\rm t} \rightarrow 0^+} z_{\rm t} \int_{z_{\rm
t}}^{\infty} z^{-1} f_{z_{\rm eff}}(z) dz=0$ when {\bf A2} is
satisfied; otherwise, based on L'H\^{o}pital's rule $\lim_{z_{\rm t}
\rightarrow 0^+} z_{\rm t} \int_{z_{\rm t}}^{\infty} z^{-1}
f_{z_{\rm eff}}(z) dz= \lim_{z_{\rm t} \rightarrow 0^+} z_{\rm t}
f_{z_{\rm eff}}(z_{\rm t})= \lim_{z_{\rm t} \rightarrow
0^+}F_{z_{\rm eff}}(z_{\rm t})- F_{z_{\rm eff}}(0)=0$. Consequently,
$\lim_{S \rightarrow \infty} d C_{\rm OA}/d (\log S)= 1-F_{z_{\rm
eff}}(0)$. Knowing that {\bf A1} holds and $z_{\rm eff}= \gamma/S$
we have $F_{z_{\rm eff}}(0)=0$. The asymptotic slope and pre-log
constant are given by
\begin{equation} \label{slope_oa_sml}
\lim_{S \rightarrow 0} \frac{d C_{\rm OA}}{d S}= \sup \{ z:
F_{z_{\rm eff}}(z) < 1 \}
\end{equation}
\begin{equation} \label{slope_oa_lg}
\hspace{-39.5mm} \lim_{S \rightarrow \infty} \frac{d C_{\rm OA}}{d
(\log S)}=1
\end{equation}
by taking the limits of (\ref{oa_slope0}) and (\ref{oa_slope1}). As
we will see from subsequent derivations, (\ref{slope_oa_sml}) and
(\ref{slope_oa_lg}) are the largest possible for any scheme in the
presence of fading, which is expected since OA is the optimal scheme
for all $S$. Typically when $z_{\rm eff}$ has infinite support,
$\lim_{S \rightarrow 0} d C_{\rm OA}/d S= \sup \{ z: F_{z_{\rm
eff}}(z) < 1 \}= \infty$ and thus OA becomes better than AWGN at low
average SNR given (\ref{slope_awgn_sml}). We will include further
discussions of this issue in Subsection \ref{asympt_sml_snr}.

\subsubsection{Truncated Channel Inversion}

For TCI, we rewrite (\ref{papr_otci}) and (\ref{cap_otci}) as
\begin{equation} \label{papr2_otci}
\hspace{-33mm} D_{\rm max}(z_{\rm t})= \left[ \int_{z_{\rm
t}}^{\infty} \frac{z_{\rm t}}{z} f_{z_{\rm eff}}(z) dz \right] ^{-1}
\end{equation}
\begin{equation} \label{cap2_otci}
C_{\rm TCI}=[1-F_{z_{\rm eff}}(z_{\rm t})] \log (1+SD_{\rm
max}(z_{\rm t}) z_{\rm t}).
\end{equation}
Equation (\ref{papr2_otci}) indicates that $D_{\rm max}$ does not
depend on $S$ given $z_{\rm t}$. By differentiating
(\ref{cap2_otci}) with respect to $S$, we obtain
\begin{equation} \label{cap_otci_diff}
\frac{dC_{\rm TCI}}{dS}= \frac{D_{\rm max} z_{\rm t}} {1+SD_{\rm
max} z_{\rm t}} [1-F_{z_{\rm eff}}(z_{\rm t})].
\end{equation}
Setting $S=0$ and substituting (\ref{papr2_otci}) in
(\ref{cap_otci_diff}), we obtain the asymptotic slope at low average
SNR
\begin{equation} \label{cap_otci_diff_p0}
\lim_{S \rightarrow 0} \frac{dC_{\rm TCI}}{dS}= z_{\rm t} D_{\rm
max} [1-F_{z_{\rm eff}}(z_{\rm t})]= \frac{1-F_{z_{\rm eff}}(z_{\rm
t})} {\int_{z_{\rm t}}^{\infty} \frac{1}{z} f_{z_{\rm eff}}(z) dz}
\end{equation}
which can be verified to be monotonically increasing with $z_{\rm
t}$ in the support of $f_{z_{\rm eff}}(z)$ and has a maximum value
of $\sup \{ z: F_{z_{\rm eff}}(z) < 1 \}$. At high average SNR we
have
\begin{equation} \label{slope_otci}
\lim_{S \rightarrow \infty} \frac{d C_{\rm TCI}} {d (\log S)}=
\lim_{S \rightarrow \infty} \frac{S d C_{\rm TCI}}{d S}= 1-F_{z_{\rm
eff}}(z_{\rm t})
\end{equation}
which is monotonically decreasing with $z_{\rm t}$ in the support of
$f_{z_{\rm eff}}(z)$.

Recall that (\ref{cap_otci_diff_p0}) and (\ref{slope_otci}) are
derived assuming $z_{\rm t}$ to be a fixed threshold, independent of
$S$. This is unlike our discussion of the OA scheme where $z_{\rm
t}$ is optimized for each $S$. Such an optimization can also be
adopted in this TCI context by maximizing (\ref{cap2_otci}) with
respect to $z_{\rm t}$. To distinguish it from a fixed threshold we
will denote the maximizer of (\ref{cap2_otci}) by $z_{\rm t}^*(S)$.
Even though it is not possible to express $z_{\rm t}^*(S)$ in closed
form, one can still analyze asymptotic expressions for a TCI scheme
with optimal threshold. In the case of the pre-log constant, it is
clear that the optimized TCI should outperform any TCI scheme with
fixed $z_{\rm t}$. From (\ref{slope_otci}) we see that the pre-log
constant of the optimized TCI should be at least $\sup_{z_{\rm t}}
(1-F_{z_{\rm eff}}(z_{\rm t}))= 1$ when {\bf A1} holds which is the
best achievable pre-log constant over fading channels in
(\ref{slope_oa_lg}). Similarly, $z_{\rm t}^*(S)$ should approach
$\sup \{ z: F_{z_{\rm eff}}(z) < 1 \}$ as $S \rightarrow 0$ since
(\ref{cap_otci_diff_p0}) is monotonically increasing with $z_{\rm
t}$.

\subsubsection{Continuous-power Truncated Channel Inversion}

We next consider the case of CTCI which unifies RA (with $z_{\rm
t}=\infty$) and CI (with $z_{\rm t}=0$). Equations (\ref{papr_ctci})
and (\ref{cap_ctci}) can be rewritten as
\begin{equation} \label{papr2_ctci}
\hspace{-62.5mm} D_{\rm max}(z_{\rm t})= \left[ F_{z_{\rm
eff}}(z_{\rm t})+ \int_{z_{\rm t}}^{\infty} \frac{z_{\rm t}}{z}
f_{z_{\rm eff}}(z) dz \right] ^{-1}
\end{equation}
\begin{equation} \label{cap2_ctci}
C_{\rm CTCI}= \int_0^{z_{\rm t}} \log (1+SD_{\rm max}z) f_{z_{\rm
eff}}(z) dz+ [1-F_{z_{\rm eff}}(z_{\rm t})] \log (1+SD_{\rm max}
z_{\rm t}).
\end{equation}
Equation (\ref{papr2_ctci}) indicates that $D_{\rm max}$ does not
depend on $S$ given $z_{\rm t}$. By differentiating
(\ref{cap2_ctci}) with respect to $S$, we obtain
\begin{equation} \label{cap_ctci_diff}
\frac{dC_{\rm CTCI}}{dS}= \int_0^{z_{\rm t}} \frac{D_{\rm max}z}
{1+SD_{\rm max}z} f_{z_{\rm eff}}(z) dz+ \frac{D_{\rm max} z_{\rm
t}} {1+SD_{\rm max} z_{\rm t}} [1-F_{z_{\rm eff}}(z_{\rm t})].
\end{equation}
Setting $S=0$ and substituting (\ref{papr2_ctci}) in
(\ref{cap_ctci_diff}), we obtain the asymptotic slope at low average
SNR $\lim_{S \rightarrow 0} dC_{\rm CTCI}/dS= (\int_0^{z_{\rm t}}
zf_{z_{\rm eff}}(z) dz+ z_{\rm t} [1-F_{z_{\rm eff}}(z_{\rm t})])/
(F_{z_{\rm eff}}(z_{\rm t})+ \int_{z_{\rm t}}^{\infty} z_{\rm t}
z^{-1} f_{z_{\rm eff}}(z) dz)$, where we exchange the limit and
integral in the first term, since the absolute value of the
integrand is upper bounded by $D_{\rm max}z f_{z_{\rm eff}}(z)$
which is an integrable function, and the condition for dominated
convergence is satisfied. Notice that this general formula for
$\lim_{S \rightarrow 0} d C_{\rm CTCI}/ d S$ is suitable for $0<
z_{\rm t}< \infty$ only, while for the cases of RA ($z_{\rm
t}=\infty$) and CI ($z_{\rm t}=0$), we can simply rewrite
(\ref{cap_cpra}) and (\ref{cap_ci}) as
\begin{equation} \label{cap2_cpra}
\hspace{10mm} C_{\rm RA}=\int_0^{\infty} \log (1+Sz) f_{z_{\rm
eff}}(z) dz
\end{equation}
\begin{equation} \label{cap2_ci}
C_{\rm CI}=\log (1+S\textrm{E}^{-1} [z_{\rm eff}^{-1}])
\end{equation}
and obtain
\begin{equation} \label{slope_cpra}
\hspace{-7mm} \lim_{S \rightarrow 0} \frac{dC_{\rm RA}}{dS}=
\textrm{E} [z_{\rm eff}]
\end{equation}
\begin{equation} \label{slope_ci}
\lim_{S \rightarrow 0} \frac{dC_{\rm CI}}{dS}= \textrm{E}^{-1}
[z_{\rm eff}^{-1}].
\end{equation}

For the analysis at high average SNR, based on (\ref{cap_ctci_diff})
we obtain
\begin{equation} \label{slope_ctci}
\lim_{S \rightarrow \infty} \frac{d C_{\rm CTCI}} {d (\log S)}=
\lim_{S \rightarrow \infty} \frac{S d C_{\rm CTCI}}{d S}=
\int_0^{z_{\rm t}} f_{z_{\rm eff}}(z) dz+ 1-F_{z_{\rm eff}}(z_{\rm
t})=1
\end{equation}
which is independent of the threshold. Here we exchange the limit
and integral since the integrand is upper bounded by an integrable
function, $f_{z_{\rm eff}}(z)$, and the condition for the dominated
convergence theorem is satisfied. Notice that (\ref{slope_ctci})
also applies to RA and CI since they are special cases of CTCI. By
comparing (\ref{slope_otci}) and (\ref{slope_ctci}), we notice that
TCI has a smaller pre-log constant than CTCI at high average SNR and
thus has a worse capacity in that regime. We will also give their
capacity comparison at low average SNR in Subsection
\ref{asympt_sml_snr}, where the opposite will be seen to hold.

Similar to OA and TCI one can seek to optimize $z_{\rm t}$ in
(\ref{cap2_ctci}) for each $S$ instead of using a fixed $z_{\rm t}$.
Interestingly, unlike the cases of OA and TCI, maximizing
(\ref{cap2_ctci}) over $z_{\rm t}$ yields the trivial answer of
$z_{\rm t}=\infty$, as stated in the following theorem.
\begin{thry} \label{ctci_thm}
$C_{\rm CTCI}$ as given by (\ref{cap2_ctci}) is monotonically
increasing with $z_{\rm t}$ for any value of $S$.
\end{thry}
\begin{proof}
See Appendix \ref{ctci_papr_mono}.
\end{proof}

Since RA and CI are obtained as special cases of CTCI when $z_{\rm
t}=\infty$ and $z_{\rm t}=0$ respectively, Theorem \ref{ctci_thm}
indicates that $C_{\rm CI} \leq C_{\rm CTCI} \leq C_{\rm RA}$
regardless of average SNR.

\subsection{Effect of Outage on the Pre-log Constant}

Recall that TCI involves outage while CTCI does not. The pre-log
constant of CTCI is $1$ whereas that of TCI is strictly less than
$1$. Without being restricted to the parametric forms of $D(z)$ in
(\ref{otci_pinst}) and (\ref{ctci_pinst}), we now generalize this
result and show that the presence of outage reduces the pre-log
constant for any instantaneous power ratio $D(z)$.
\begin{thry} \label{slope1_thm}
Let $D(z)$ be independent of $S$, define $\mathscr{O}= \{ z| D(z)=0
\}$, and assume {\bf A1} is satisfied. Then the pre-log constant of
the scheme with instantaneous power ratio $D(z)$ is given by
\begin{equation}
\lim_{S \rightarrow \infty} \frac{d \textrm{E}[\log (1+S D(z_{\rm
eff}) z_{\rm eff})]}{d (\log S)}= 1-\textrm{Pr}(\mathscr{O}).
\end{equation}
\end{thry}
\begin{proof}
See Appendix \ref{slope1_condt}.
\end{proof}

Theorem \ref{slope1_thm} clarifies that the pre-log constant being
$1$ or strictly less than $1$ is related to the absence or presence
of outage. Applying Theorem \ref{slope1_thm}, it is easy to see that
the pre-log constants of RA, CI and CTCI are $1$, provided that {\bf
A1} is satisfied. In contrast, the pre-log constant of TCI is given
by $1-F_{z_{\rm eff}}(z_{\rm t})$. Therefore, we expect constant
capacity difference gaps among the capacity curves of AWGN, OA, RA,
CI and CTCI at sufficiently high average SNR. In what follows, we
use the term ``gap'' to denote $\lim_{S \rightarrow \infty} (C_{\rm
1}-C_{\rm 2})$ for schemes 1 and 2 satisfying $\lim_{S \rightarrow
\infty} d C_{\rm 1}/d (\log S)=\lim_{S \rightarrow \infty} d C_{\rm
2}/d (\log S)=1$. The corresponding gap in average SNR in dB can be
derived to be simply $10/\log 10$ times the capacity difference gap
above.

\subsection{Asymptotic Gaps among AWGN, OA, RA and CI at High Average
SNR} \label{asympt_gaps}

To derive the high-SNR asymptotic capacity gaps, we introduce the
following.
\begin{lem} \label{lem_int}
For a constant $\mu>0$ and a non-negative random variable $z_{\rm
eff}$ having finite E$[\log z_{\rm eff}]$, we have
\begin{equation} \label{lem_int_eqn1}
\lim_{S \rightarrow \infty} \int_{0}^{\infty} \log \left(
\frac{1+S\mu}{S\mu} \right) f_{z_{\rm eff}}(z) dz=0
\end{equation}
\begin{equation} \label{lem_int_eqn2}
\lim_{S \rightarrow \infty} \int_{0}^{\infty} \log \left(
\frac{1+Sz}{Sz} \right) f_{z_{\rm eff}}(z) dz= 0
\end{equation}
\end{lem}
\begin{proof}
See Appendix \ref{lem_int_proof}.
\end{proof}
Notice that the assumption $-\infty< {\rm E}[\log z_{\rm eff}]<
\infty$ we have in Lemma \ref{lem_int} follows from {\bf A1} and is
weaker than {\bf A2}, while {\bf A2} can also be assumed to prove
Lemma \ref{lem_int}. Consequently we have the following theorem
\begin{thry} \label{th_gaps}
The following high-SNR capacity difference gaps among the OA, RA, CI
schemes and AWGN capacity are given by
\begin{equation} \label{gap_oa_ra}
\lim_{S \rightarrow \infty} (C_{\rm OA}-C_{\rm RA})=0
\end{equation}
\begin{equation} \label{gap_awgn_oa}
\lim_{S \rightarrow \infty} (C_{\rm AWGN}-C_{\rm OA})= \log
(\textrm{E}[z_{\rm eff}])- \textrm{E}[\log z_{\rm eff}] \geq 0
\end{equation}
\begin{equation} \label{gap_oa_ci}
\lim_{S \rightarrow \infty} (C_{\rm OA}-C_{\rm CI})= \textrm{E}[\log
z_{\rm eff}]+ \log (\textrm{E}[z_{\rm eff}^{-1}]) \geq 0
\end{equation}
Moreover, (\ref{gap_awgn_oa}) is finite when {\bf A1} holds, and
(\ref{gap_oa_ci}) is finite when {\bf A2} holds.
\end{thry}
\begin{proof}
See Appendix \ref{th_gaps_proof}.
\end{proof}

Theorem \ref{th_gaps} indicates that the AWGN capacity exceeds the
optimal capacity under any fading channel distribution at high SNR.
However, in \ref{asympt_sml_snr} we will show the exact opposite at
low SNR. We have thus established that OA and RA exhibit zero gap at
high average SNR, and that (\ref{gap_awgn_oa}) yields the
(non-negative) gap between AWGN and OA. We recall that as a
consequence of the discussion after Theorem \ref{ctci_thm},
\begin{equation}
C_{\rm CI} \leq C_{\rm CTCI} \leq C_{\rm RA} \leq C_{\rm OA}
\end{equation}
for any $S$, and therefore the gap between $C_{\rm OA}$ and $C_{\rm
CI}$, given by (\ref{gap_oa_ci}), is the largest constant gap among
all the schemes over fading channels we have addressed.

\subsection{Asymptotic Comparisons at Low Average SNR} \label{asympt_sml_snr}

From the asymptotic analysis at high average SNR, we recall that the
presence of outage in adaptive transmission schemes tends to result
in worse capacity than no outage (Theorem \ref{slope1_thm}) and AWGN
capacity is better than all the capacities under fading based on
(\ref{gap_awgn_oa}). However, capacity comparisons at low average
SNR are sharply different. We have the following results for
selected schemes at low average SNR.
\begin{thry} \label{th_lowsnr}
The low-SNR slopes of capacities satisfy the following:
\begin{equation} \label{lowsnr_comp1}
\textrm{CTCI: } \textrm{E}^{-1} [z_{\rm eff}^{-1}]= \lim_{S
\rightarrow 0} \frac{dC_{\rm CI}}{dS} \leq \lim_{S \rightarrow 0}
\frac{dC_{\rm CTCI}}{dS} \leq \lim_{S \rightarrow 0} \frac{dC_{\rm
RA}}{dS}= \textrm{E} [z_{\rm eff}]
\end{equation}
\begin{equation} \label{lowsnr_comp3}
\hspace{0.5mm} \textrm{OA: } \hspace{3mm} \lim_{S \rightarrow 0}
\frac{dC_{\rm OA}}{dS}= \sup\{ z: F_{z_{\rm eff}}(z) < 1 \} \geq
\textrm{E} [z_{\rm eff}]= \lim_{S \rightarrow 0} \frac{dC_{\rm
AWGN}}{dS}
\end{equation}
\begin{equation} \label{lowsnr_comp2}
\hspace{-20mm} \textrm{TCI: } \hspace{3mm} \textrm{E}^{-1} [z_{\rm
eff}^{-1}] \leq \lim_{S \rightarrow 0} \frac{dC_{\rm TCI}}{dS} \leq
\sup \{ z: F_{z_{\rm eff}}(z) < 1 \},
\end{equation}
where, in (\ref{lowsnr_comp1}) the two inequalities become
equalities when the threshold satisfies $z_{\rm t}=0$ and $\infty$
respectively; in (\ref{lowsnr_comp3}) the equality holds only when
the fading channel is deterministic and reduces to AWGN; in
(\ref{lowsnr_comp2}) the two inequalities become equalities when the
threshold $z_{\rm t}=0$ and $\sup \{ z: F_{z_{\rm eff}}(z) < 1 \}$
respectively. Also for TCI with a threshold $z_{\rm t}$, $z_{\rm t}
\geq \textrm{E} [z_{\rm eff}]$ is a sufficient condition for
$\lim_{S \rightarrow 0} dC_{\rm TCI}/dS \geq \lim_{S \rightarrow 0}
dC_{\rm AWGN}/dS$.
\end{thry}
\begin{proof}
See Appendix \ref{th_lowsnr_proof}.
\end{proof}

Theorem \ref{th_lowsnr} indicates that OA gives larger capacity than
AWGN at low SNRs for {\it any} fading distribution, and the presence
of fading {\it always} improves capacity when the average SNR is
sufficiently low. To the best of our knowledge, this is proved
analytically here for the first time, even though it has been
briefly mentioned in \cite{goldsmith97} and addressed in
\cite{alouini00} with a numerical example for specific distributions
without an analytical proof and addressed in \cite{tsebook05}
through approximations. Also, with the choice $z_{\rm t} \geq
\textrm{E} [z_{\rm eff}]$, TCI can also give larger capacity than
AWGN at low average SNRs, which to our knowledge was never mentioned
in existing literature. It follows that with a sufficiently low
average SNR, the presence of fading can result in better capacity
than the equivalent AWGN channel. Furthermore, the comparison
between (\ref{lowsnr_comp1}) and (\ref{lowsnr_comp2}) suggests that
outage can be helpful to improve the low-SNR capacity by exploiting
the aforementioned benefit of fading.

\section{Asymptotic Optimality of CI with Diversity} \label{gap_div}

It is clear that the presence of diversity will improve the ergodic
capacity of all the schemes discussed. However, the high-SNR
capacity gap between OA and the sub-optimal CI also reduces in the
presence of diversity rendering CI near-optimal in some cases. In
this section, we give examples of space diversity and multi-user
diversity to show that the gap given by (\ref{gap_oa_ci}) between CI
and OA can be made arbitrarily small with sufficiently large number
of antennas or users. Since (\ref{gap_awgn_oa}) is non-negative,
$\lim_{S \rightarrow \infty} (C_{\rm AWGN}-C_{\rm CI}) \geq \lim_{S
\rightarrow \infty} (C_{\rm OA}-C_{\rm CI})$, it will also be
convenient to investigate
\begin{equation} \label{gap_awgn_ci}
\lim_{S \rightarrow \infty} (C_{\rm AWGN}-C_{\rm CI})= \lim_{S
\rightarrow \infty} (C_{\rm AWGN}-C_{\rm OA})+ \lim_{S \rightarrow
\infty} (C_{\rm OA}-C_{\rm CI})= \log (\textrm{E}[z_{\rm eff}]
\textrm{E}[z_{\rm eff}^{-1}])
\end{equation}
in the presence of diversity. Clearly, showing that the gap in
(\ref{gap_awgn_ci}) can be made arbitrarily small in the presence of
diversity would establish the same for (\ref{gap_oa_ci}).

\subsection{Example of Space Diversity}

For space diversity, we consider the case in which the system
consists of a transmitter with $N \geq 2$ antennas and a receiver
with a single antenna, and the instantaneous SNR is to be maximized
through beamforming. We assume the components $\{ h_n \}_{n=1}^N$ of
the $N \times 1$ channel vector are independently and identically
distributed (i.i.d.) circularly symmetric $\mathcal{CN}(0,1)$ random
variables, and the noise term is also circularly symmetric
$\mathcal{CN}(0,1)$. It is well-known that the maximized effective
channel gain is $z_{\rm eff}=\sum_{n=1}^N |h_n|^2$, whose PDF is
given by $f_{z_{\rm eff}}(z)=z^{N-1} e^{-z}/ \Gamma(N)$. Here
$\Gamma(x):= \int_0^{\infty} t^{x-1} e^{-t} dt$ is the Gamma
function and for integer $N$ it becomes $\Gamma(N)= (N-1)!$. It can
easily be verified that {\bf A1} is satisfied with $d=N$. The gaps
in (\ref{gap_oa_ci}) and (\ref{gap_awgn_ci}) become
\begin{equation} \label{gap_oa_ci_n}
\begin{split}
\lim_{S \rightarrow \infty} (C_{\rm OA}-C_{\rm CI}) &=
\int_0^{\infty} \log z f_{z_{\rm eff}}(z) dz+ \log
\left( \int_0^{\infty} \frac{1}{z} f_{z_{\rm eff}}(z) dz \right)\\
&= \psi(N)- \log (N-1)
\end{split}
\end{equation}
\begin{equation} \label{gap_awgn_ci_n}
\begin{split}
\hspace{1mm} \lim_{S \rightarrow \infty} (C_{\rm AWGN}-C_{\rm CI})
&= \log \left( \int_0^{\infty} zf_{z_{\rm eff}}(z) dz \right)+ \log
\left( \int_0^{\infty} \frac{1}{z} f_{z_{\rm eff}}(z)
dz \right)\\
&= \log N+ \log \left( \frac{1}{N-1} \right)= \log \left(
1+\frac{1}{N-1} \right)
\end{split}
\end{equation}
where $\psi(x)=d \log \Gamma(x)/d x$ is the digamma function. In
\cite{biglieri98} $C_{\rm CI}$ was derived for $N>1$ and termed
delay-limited capacity, and is given by $\log (1+ (N-1)S/N)$. The
result in (\ref{gap_awgn_ci_n}) can also be obtained using the
expression for $C_{\rm CI}$ in \cite{biglieri98}. However, the gap
in (\ref{gap_oa_ci_n}) was never addressed therein. We observe from
(\ref{gap_oa_ci_n}) and (\ref{gap_awgn_ci_n}) that the gaps can be
made arbitrarily small for sufficiently large number of antennas.
Therefore, CI provides near-optimal capacity at high average SNR
with space diversity. Moreover, for large $N$ we have
\begin{equation} \label{gap_oa_ci_largen}
\psi(N)- \log (N-1)= \frac{1}{2(N-1)}+ \mathcal{O} \left( N^{-2}
\right)
\end{equation}
\begin{equation} \label{gap_awgn_ci_largen}
\log \left( 1+ \frac{1}{N-1} \right)= \frac{1}{(N-1)}+ \mathcal{O}
\left( N^{-2} \right) .
\end{equation}
It can be seen that (\ref{gap_oa_ci_n}) is approximately half of
(\ref{gap_awgn_ci_n}) and they both behave inversely proportional to
$N-1$ with $N$ being sufficiently large. We note that a similar
result can be obtained for receive diversity instead of transmit
beamforming.

\subsection{Example of Multi-user Diversity}

For multi-user diversity, we consider the case in which the
transmitter selects the user (receiver) among $K \geq 2$ users with
the best channel for transmission at any given time instant.
Specifically, the effective channel gain becomes $z_{\rm
eff}=\max_{k=1,2,...,K} \{ z_k \}$ where $z_k$ is the individual
channel gain when the $k$-th user is selected for transmission. In
general, the problem is more complicated than space diversity since
we are unable to get closed-form results of the gaps in
(\ref{gap_oa_ci}) and (\ref{gap_awgn_ci}) for the commonly
considered wireless channels. However, as we now see it is possible
to quantify how the gap given by (\ref{gap_awgn_ci}) scales with the
number of users $K$.

We investigate the case in which $z_k$ has exponential CDF
$F(z)=1-e^{-z}$, corresponding to a single-antenna system under
Rayleigh fading with $K$ users. In this case, $z_{\rm eff}$ has CDF
$F_{z_{\rm eff}}(z)=(1-e^{-z})^K$ and PDF $f_{z_{\rm eff}}(z)=K
e^{-z} (1-e^{-z})^{K-1}$, with $d=K$ in {\bf A1}. It is well-known
that $\textrm{E}[z_{\rm eff}]=\sum_{k=1}^K 1/k$. Adapting Lemma 3.1
in \cite{leong10}, it can be shown that $\textrm{E}[z_{\rm
eff}^{-1}] \sim 1/ \log K$ (here ``$\sim$'' denotes asymptotical
equality) as $K \rightarrow \infty$. Moreover since $\lim_{K
\rightarrow \infty} (\sum_{k=1}^K 1/k- \log K)= \gamma_{\rm em}$
where $\gamma_{\rm em} \approx 0.5772$ is the Euler-Mascheroni
constant, for $K \rightarrow \infty$ the high-SNR gap between
$C_{\rm AWGN}$ and $C_{\rm CI}$ given by (\ref{gap_awgn_ci}) becomes
\begin{equation} \label{gap_awgn_ci_k}
\log (\textrm{E}[z_{\rm eff}] \textrm{E}[z_{\rm eff}^{-1}]) \sim
\log \left( \frac{\log K+ \gamma_{\rm em}}{\log K} \right)= \log
\left( 1+ \frac{\gamma_{\rm em}}{\log K} \right) \sim
\frac{\gamma_{\rm em}}{\log K} .
\end{equation}
Due to the limitation of mathematical tools, we are unable to derive
any asymptotic expression of $\lim_{S \rightarrow \infty} (C_{\rm
OA}-C_{\rm CI})$ rigorously under multi-user diversity, however we
conjecture $\lim_{S \rightarrow \infty} (C_{\rm OA}-C_{\rm CI})=
{\rm E}[\log z_{\rm eff}]+ \log ({\rm E}[z_{\rm eff}^{-1}]) \sim
\gamma_{\rm em}/(\log K (1+\log K))$ for sufficiently large $K$
based on a non-rigorous approach, which evaluates ${\rm E}[\log
z_{\rm eff}]$ and $\log ({\rm E}[z_{\rm eff}^{-1}])$ directly based
on the extreme distributions of $\log z_{\rm eff}$ and $z_{\rm
eff}^{-1}$ without verifying the condition of uniform integrability
\cite[ch.6]{dasguptabook08}. Clearly (\ref{gap_awgn_ci_k}) converges
very slowly to zero as the number of users increases, and
consequently so does the gap between OA and CI as per the discussion
after (\ref{gap_awgn_ci}). Therefore, multi-user diversity can also
render CI achieve near-AWGN capacity, but the number of users needed
to close the gap given by (\ref{gap_awgn_ci_k}) is much larger than
the number of antennas needed for (\ref{gap_awgn_ci_n}).

It is interesting to note that for the multi-user diversity example,
(\ref{gap_oa_ci}) and (\ref{gap_awgn_ci}) do not always converge to
zero as $K \rightarrow \infty$ if the SNR distribution of a single
user is not exponential. For example, when $z_k$ is of Fr\'{e}chet
distribution with CDF $F(z)=\exp(-z^{-\alpha})$, it can be shown
that (\ref{gap_oa_ci}) and (\ref{gap_awgn_ci}) are given by
\begin{equation}
\textrm{E}[\log z_{\rm eff}]+ \log (\textrm{E}[z_{\rm eff}^{-1}])=
\frac{\gamma_{\rm em}} {\alpha}+ \log \Gamma \left(
1+\frac{1}{\alpha} \right)
\end{equation}
\begin{equation}
\log (\textrm{E}[z_{\rm eff}] \textrm{E}[z_{\rm eff}^{-1}])= \log
\left( \Gamma \left( 1-\frac{1}{\alpha} \right) \Gamma \left(
1+\frac{1}{\alpha} \right) \right)
\end{equation}
which do not depend on $K$! In such cases, it becomes impossible for
CI to achieve the optimal or AWGN capacity by simply increasing the
number of users. Fr\'{e}chet distribution is known to arise from the
extreme signal-to-interference ratio distribution in
interference-limited MIMO wireless channels with sufficiently large
number of users \cite{pun07}.

\section{Simulation Results} \label{simres}

In this section, we provide simulation results to corroborate our
theoretical analysis. In all simulations we model the system as a
multi-antenna and/or multi-user MISO system with $N$ antennas at the
base station (transmitter) and a single antenna at each of the $K$
users (receivers). Only the user with the best channel condition is
selected for transmission at each time instant. Specifically, we
assume the channel coefficients between antenna $n$ and user $k$,
$h_{nk}$, are i.i.d. circularly symmetric $\mathcal{CN}(0,1)$ random
variables and so are the noise terms. Then the maximized effective
channel gain becomes $z_{\rm eff}= \max_{k=1,2,...,K} \{
\sum_{n=1}^N |h_{nk}|^2 \}$, whose PDF is given by $f_{z_{\rm
eff}}(z)= K (\Gamma(N))^{-K} [\gamma (N,z)]^{K-1} z^{N-1} e^{-z}$.
It can be proved that {\bf A2} holds if and only if $\max (N,K) \geq
2$, i.e. at least one kind of diversity is present. Recall that the
average SNR is the average power $S$ in linear scale since the
channel noise has normalized variance. In the plots, we exhibit our
results using base-2 logarithm instead of natural logarithm, so that
all the capacity values and gaps are in bits per channel use.

\begin{figure}[!ht]
\begin{minipage}{1.0\textwidth}
\begin{center}
\includegraphics[height=9cm,keepaspectratio]{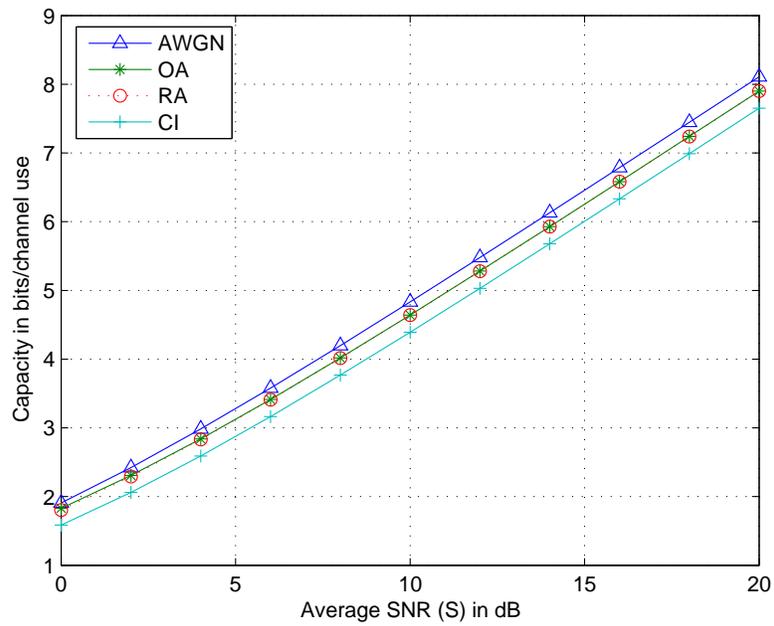}
\caption{Capacities of AWGN, OA, RA and CI for $(N,K)=(2,2)$}
\label{cap_awgn_oa_cpra_ci}
\end{center}
\end{minipage}
\end{figure}

\begin{figure}[!ht]
\begin{minipage}{1.0\textwidth}
\begin{center}
\includegraphics[height=9cm,keepaspectratio]{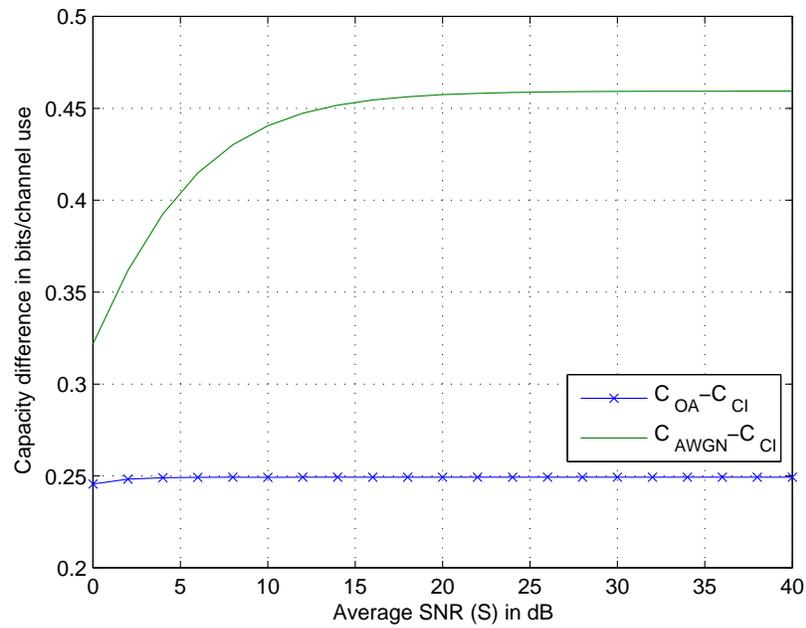}
\caption{Capacity differences from OA and AWGN to CI at different
average SNRs for $(N,K)=(2,2)$} \label{gap_ci_snr}
\end{center}
\end{minipage}
\end{figure}

Figure \ref{cap_awgn_oa_cpra_ci} shows the capacity curves of OA,
RA, CI and AWGN for $(N,K)=(2,2)$. We observe that RA achieves
almost the same capacity as OA. In addition, all three schemes seem
to have the same pre-log constant as the AWGN capacity at high
average SNR. All these observations corroborate our analysis in
Section \ref{asympt}. Figure \ref{gap_ci_snr} shows the capacity
differences at finite average SNRs. We observe that for average SNR
higher than $6$ dB, the gap between OA and CI becomes almost steady
at the value $0.24928$ (bits) also predicted by our analytical
results in (\ref{gap_oa_ci}). However, the value $0.45943$ obtained
through (\ref{gap_awgn_ci}) which is the asymptotic gap between AWGN
and CI requires significantly higher average SNR (about $40$ dB) to
become visible.

\begin{figure}[!ht]
\begin{minipage}{1.0\textwidth}
\begin{center}
\includegraphics[height=9cm,keepaspectratio]{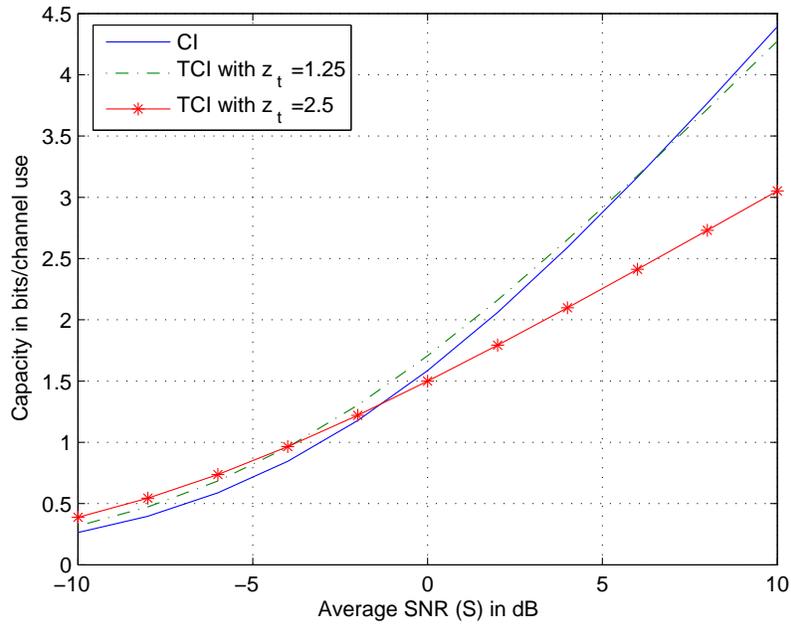}
\caption{Capacities of CI and TCI for $(N,K)=(2,2)$}
\label{cap_ci_otci}
\end{center}
\end{minipage}
\end{figure}

\begin{figure}[!ht]
\begin{minipage}{1.0\textwidth}
\begin{center}
\includegraphics[height=9cm,keepaspectratio]{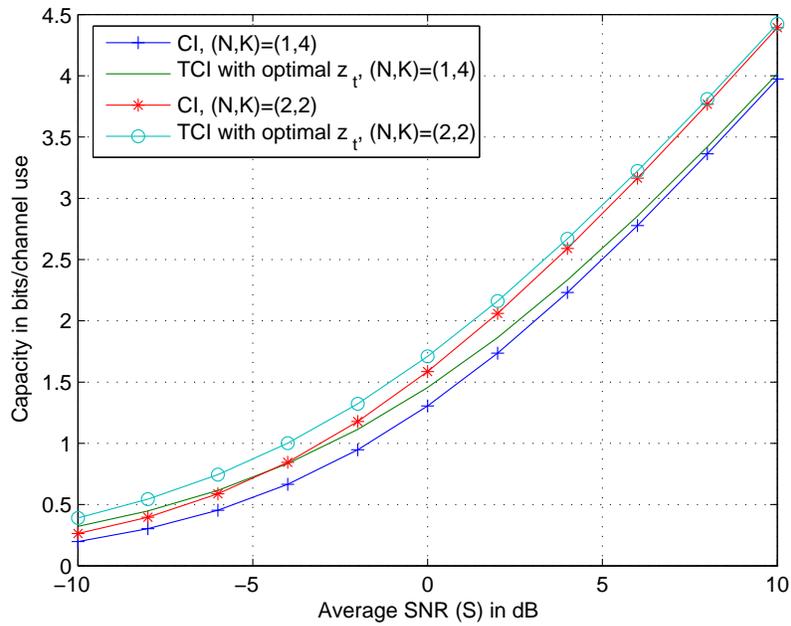}
\caption{Capacities of CI and optimized TCI for $(N,K)=(1,4)$ and
$(2,2)$} \label{cap_ci_otci_opt}
\end{center}
\end{minipage}
\end{figure}

\begin{figure}[!ht]
\begin{minipage}{1.0\textwidth}
\begin{center}
\includegraphics[height=9cm,keepaspectratio]{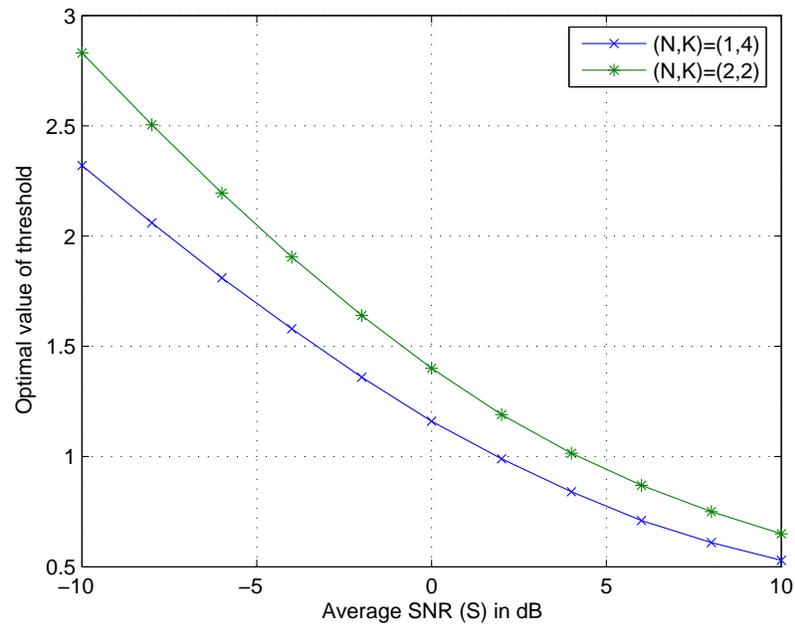}
\caption{Optimal $z_{\rm t}$ of TCI for $(N,K)=(1,4)$ and $(2,2)$}
\label{opt_th}
\end{center}
\end{minipage}
\end{figure}

Figure \ref{cap_ci_otci} shows the capacity curves of two TCI
schemes with different fixed thresholds as well as CI for the case
of $(N,K)=(2,2)$. We observe that TCI exhibits a smaller pre-log
constant than CI at high average SNR, and a larger threshold results
in a smaller pre-log constant. Consequently, TCI has worse capacity
than CI at high average SNR, as expected. However, the conclusion is
reversed at sufficiently low average SNR: TCI gives better capacity
than CI, and a larger threshold is more favorable. This corroborates
our analysis in Subsections \ref{asympt_slopes} and
\ref{asympt_sml_snr}. In Figure \ref{cap_ci_otci_opt} we compare TCI
with optimal threshold and CI for $(N,K)=(1,4)$ and $(2,2)$. Figure
\ref{opt_th} shows the optimal thresholds of TCI at different
average SNRs for $(N,K)=(1,4)$ and $(2,2)$. We observe that as the
average SNR increases, the optimal threshold of TCI becomes smaller
from Figure \ref{opt_th}. Consequently, the capacity of TCI becomes
closer to the capacity of CI at higher average SNR in both choices
of $(N,K)$, as can be observed from Figure \ref{cap_ci_otci_opt}.
These observations corroborate our analysis in \ref{asympt_slopes}.

\begin{figure}[!ht]
\begin{minipage}{1.0\textwidth}
\begin{center}
\includegraphics[height=9cm,keepaspectratio]{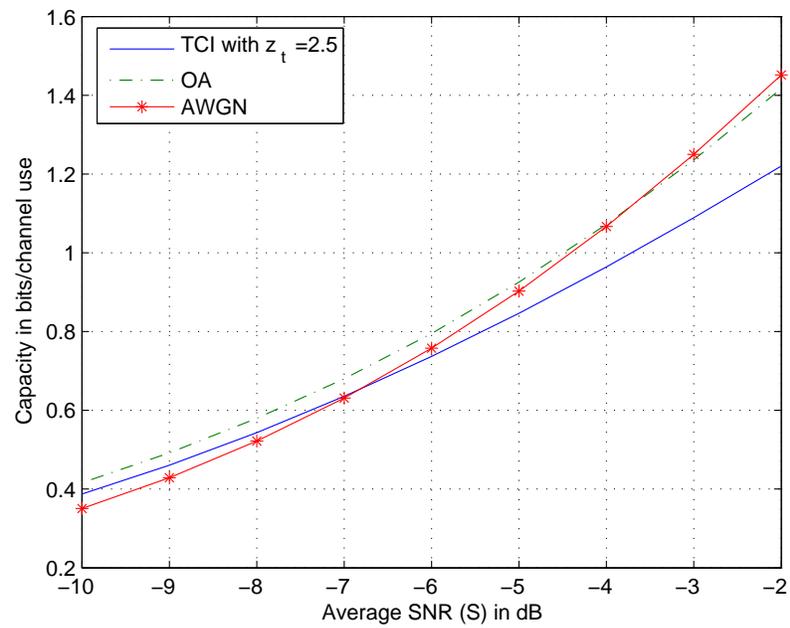}
\caption{Capacities of TCI, OA and AWGN for $(N,K)=(2,2)$ at low
average SNR} \label{cap_otci_smlsnr}
\end{center}
\end{minipage}
\end{figure}

In Figure \ref{cap_otci_smlsnr} we give comparisons of TCI, OA and
AWGN, at low average SNR. We observe that both OA and TCI can
provide better capacity than the AWGN case, which verifies our
analysis in Subsection \ref{asympt_sml_snr}. In addition, we
mentioned in Subsection \ref{asympt_sml_snr} that $z_{\rm t} \geq
\textrm{E}[z_{\rm eff}]$ is sufficient for TCI to have better
capacity than AWGN. However for $(N,K)=(2,2)$ we have
$\textrm{E}[z_{\rm eff}]=2.75>2.5$, which indicates that $z_{\rm t}
\geq \textrm{E}[z_{\rm eff}]$ is not actually necessary. Since TCI
has significantly less implementation complexity than OA, it is
shown to be a viable adaptive transmission scheme at low average
SNR.

\begin{figure}[!ht]
\begin{minipage}{1.0\textwidth}
\begin{center}
\includegraphics[height=9cm,keepaspectratio]{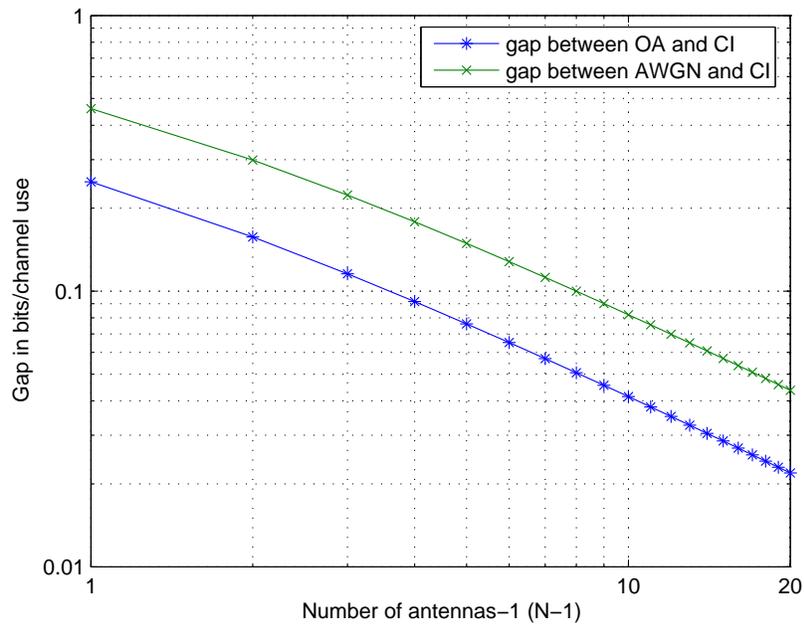}
\caption{$\lim_{S \rightarrow \infty} (C_{\rm OA}-C_{\rm CI})$ and
$\lim_{S \rightarrow \infty} (C_{\rm AWGN}-C_{\rm CI})$ with
transmit or receive diversity between two points} \label{gap_ci_n}
\end{center}
\end{minipage}
\end{figure}

\begin{figure}[!ht]
\begin{minipage}{1.0\textwidth}
\begin{center}
\includegraphics[height=9cm,keepaspectratio]{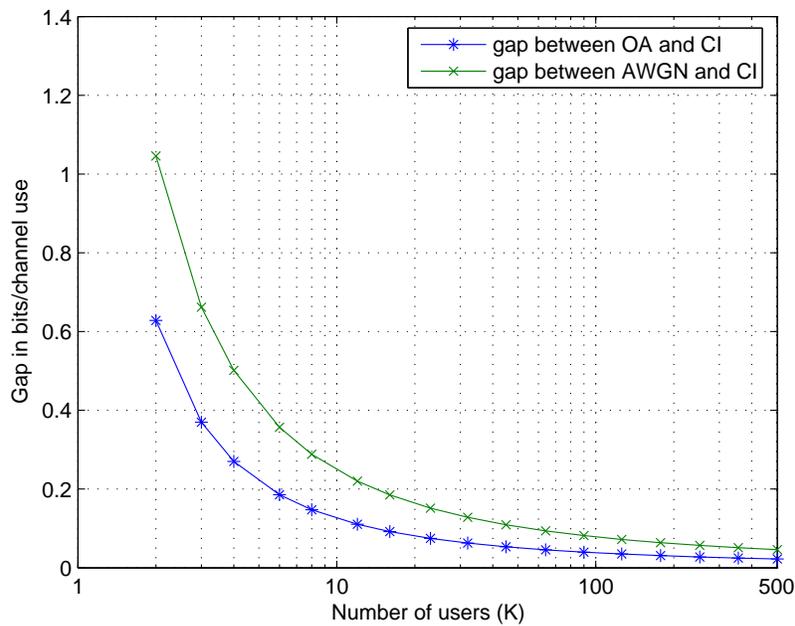}
\caption{$\lim_{S \rightarrow \infty} (C_{\rm OA}-C_{\rm CI})$ and
$\lim_{S \rightarrow \infty} (C_{\rm AWGN}-C_{\rm CI})$ with single
transmit and receive antennas and multi-user diversity}
\label{gap_ci_k}
\end{center}
\end{minipage}
\end{figure}

Figure \ref{gap_ci_n} shows the asymptotic gap between OA and CI
obtained from (\ref{gap_oa_ci}) and the asymptotic gap between AWGN
and CI obtained from (\ref{gap_awgn_ci}), for $K=1$ and different
values of $N$. We observe that the gaps are both decreasing and
converging to zero with the increase of diversity order, and appear
to be inversely proportional to $N-1$ since the plots become almost
straight lines with slope $-1$, as suggested by
(\ref{gap_oa_ci_largen}) and (\ref{gap_awgn_ci_largen}). Figure
\ref{gap_ci_k} shows the same thing depicted by Figure
\ref{gap_ci_n} for $N=1$ and different values of $K$. It also
displays the fact that the gaps become smaller and converge to zero
as the number of users increases, but the decrease is much slower
than the case of Figure \ref{gap_ci_n} since the gap between AWGN
and CI is inversely proportional to $\log K$.

\section{Conclusions} \label{concl}

In this paper we investigate asymptotic properties at both high and
low average SNRs of the ergodic capacities for several adaptive
transmission schemes and a wide class of channel distributions. We
show that at high average SNR, both CI and CTCI exhibit the same
capacity pre-log constant of $1$, while TCI exhibits a pre-log
constant which is strictly less than $1$. This is a special case of
a more general result (Theorem \ref{slope1_thm}) which shows that
the presence of outage in the high-SNR instantaneous power ratio
reduces the pre-log constant. Consequently, with the average SNR
being sufficiently high, both CI and CTCI outperform TCI. In
addition, we prove that the capacity of CTCI is monotonically
increasing with the threshold for any value of average SNR, which is
used to show that the largest asymptotic capacity gap among all the
schemes under fading is the one between CI and OA. We have derived
closed-form expressions for asymptotic gaps between CI and OA as
well as CI and AWGN. For the case of low average SNR, we show that
AWGN capacity can be exceeded by OA and TCI. Consequently, TCI is a
favorable scheme at low average SNR since it has significantly less
complexity than OA.

We also study the behavior of the derived high-SNR gaps among CI, OA
and AWGN in the presence of diversity. Through the examples of space
diversity and multi-user diversity, we point out that the high-SNR
gaps can be made arbitrarily small with sufficiently large number of
antennas or users. Based on our expressions for rates of
convergence, it is shown that antennas are more efficient than users
in reducing the gaps. An example of a channel distribution under
which the sub-optimality gaps are independent of $K$ is also given
to illustrate that the gaps need not always reduce with $K$. This
indicates that under certain conditions CI is asymptotically optimal
with sufficiently large diversity order.

\appendices

\section{Implications of {\bf A1}} \label{reg_vary_proof}

In this appendix, we show that {\bf A1} implies $F_{\gamma}(0)=0$
and $|\textrm{E}[\log \gamma]|< \infty$. Since $F_{\gamma}(x)$ is
regularly varying at the origin, we have $\lim_{x \rightarrow 0}
F_{\gamma}(\tau x)/F_{\gamma}(x)= \tau^{d}$ for $\tau>0$. Clearly,
as $\tau \rightarrow 0$ it is required that
$F_{\gamma}(0)/F_{\gamma}(x)=0$ since $d>0$. Moreover since $0 \leq
F_{\gamma}(x) \leq 1$, we obtain $F_{\gamma}(0)=0$.

We next prove that E$[|\log \gamma|]< \infty$, which implies that
$|\textrm{E}[\log \gamma]|< \textrm{E}[|\log \gamma|]< \infty$.
Define $\delta= \sup \{ x: F_{\gamma}(x)< 1 \}$. We consider the
cases of $0< \delta \leq 1$ and $\delta> 1$. We will use for both
cases, since $\log x$ is slowly varying at $x=0$, $\log x
F_{\gamma}(x)$ is regularly varying with exponent $d>0$ similar to
$F_{\gamma}(x)$, thus $\lim_{x \rightarrow 0} \log x F_{\gamma}(x)=
0$.

If $0< \delta \leq 1$, we have $\textrm{E}[|\log \gamma|]=
-\int_0^{\delta} \log x f_{\gamma}(x) dx= \int_0^{\delta} x^{-1}
F_{\gamma}(x) dx- \log \delta$ using integration by parts and
$\lim_{x \rightarrow 0} \log x F_{\gamma}(x)= 0$. Let $y=x^{-1}$,
$l(x)=r(x^{-1})=r(y)$ and $F_{\gamma}(x)=G(x^{-1})=G(y)$, we have
$\int_0^{\delta} x^{-1} F_{\gamma}(x) dx= \int_{1/ \delta} ^{\infty}
y^{-1} G(y) dy= \int_{1/ \delta} ^{\infty} y^{-1-d} r(y) dy$, and
$r(y)$ varies slowly at $\infty$. Consequently, we have
$\int_0^{\delta} x^{-1} F_{\gamma}(x) dx= \int_{1/ \delta} ^{\infty}
y^{-1-d} r(y) dy< \infty$, which can be justified by a modification
of the Lemma in \cite[pp.280]{fellerbook71} given $-1-d< -1$. It
follows that $\textrm{E}[|\log \gamma|]= \int_0^{\delta} x^{-1}
F_{\gamma}(x) dx- \log \delta< \infty$.

If $\delta> 1$, similar to the previous case, we have
$\textrm{E}[|\log \gamma|]= \int_0^1 x^{-1} F_{\gamma}(x) dx+
\int_1^{\delta} \log x f_{\gamma}(x) dx$, and it can be proved that
$\int_0^1 x^{-1} F_{\gamma}(x) dx< \infty$. Furthermore, since $\log
x \leq x-1$, $\int_1^{\delta} \log x f_{\gamma}(x) dx \leq
\int_1^{\delta} (x-1) f_{\gamma}(x) dx< \textrm{E}[\gamma]+
F_{\gamma}(1)- 1< \infty$. It follows that $\textrm{E}[|\log
\gamma|]< \infty$.

\section{Proof of Lemma \ref{lem_zt}} \label{lem_zt_proof}

We first show that $\limsup_{S \rightarrow 0} z_{\rm t}=\sup \{ z:
F_{z_{\rm eff}}(z) < 1 \}$. If instead, there exists $\delta$ such
that $z_{\rm t} \leq \delta< \sup \{ z: F_{z_{\rm eff}}(z) < 1 \}$
for all $S$ then
\begin{equation} \label{oa_pave2_bound}
\frac{1}{S} \int_{\delta}^{\infty} \left( \frac{1}{\delta}-
\frac{1}{z} \right) f_{z_{\rm eff}}(z) dz \leq \int_{z_{\rm
t}}^{\infty} \frac{1}{S} \left( \frac{1}{z_{\rm t}}- \frac{1}{z}
\right) f_{z_{\rm eff}}(z) dz
\end{equation}
since the left hand side of (\ref{oa_pave2}) is monotonically
decreasing with $z_{\rm t}$ for a fixed $S$. However,
(\ref{oa_pave2_bound}) cannot hold since as $S \rightarrow 0$ it
violates (\ref{oa_pave2}). Moreover since from (\ref{diff_p_zt})
$z_{\rm t}$ is monotonically decreasing with $S$, we have $\lim_{S
\rightarrow 0} z_{\rm t}=\sup \{ z: F_{z_{\rm eff}}(z) < 1 \}$. It
can be seen from (\ref{sfn_zt}) that $0 \leq z_{\rm t}< 1/S$, and
consequently $\lim_{S \rightarrow \infty} z_{\rm t}=0$.

\section{Proof of Theorem \ref{ctci_thm}} \label{ctci_papr_mono}

Differentiating (\ref{cap2_ctci}) with respect to $z_{\rm t}$, we
obtain
\begin{equation} \label{cap_ctci_diff_zct}
\begin{split}
\frac{d C_{\rm CTCI}}{d z_{\rm t}} &= \int_0^{z_{\rm t}} \frac{-S
D_{\rm max}^2 z} {1+S D_{\rm max} z} \left( \int_{z_{\rm
t}}^{\infty} \frac{1}{z} f_{z_{\rm eff}}(z) dz \right) f_{z_{\rm
eff}}(z) dz\\
&+ \frac{1-F_{z_{\rm eff}}(z_{\rm t})} {1+S D_{\rm max} z_{\rm t}}
\left( S D_{\rm max}- S D_{\rm max}^2 z_{\rm t} \int_{z_{\rm
t}}^{\infty} \frac{1}{z} f_{z_{\rm eff}}(z) dz \right)\\
&= F_{z_{\rm eff}}(z_{\rm t}) [1-F_{z_{\rm eff}}(z_{\rm t})]
\frac{S D_{\rm max}^2} {1+S D_{\rm max} z_{\rm t}}\\
&- \left( \int_{z_{\rm t}}^{\infty} \frac{1}{z} f_{z_{\rm eff}}(z)
dz \right) \left( \int_0^{z_{\rm t}} \frac{S D_{\rm max}^2 z} {1+S
D_{\rm max} z} f_{z_{\rm eff}}(z) dz \right)
\end{split}
\end{equation}
Notice that we take $D_{\rm max}$ as a function of $z_{\rm t}$ given
by (\ref{papr2_ctci}) in (\ref{cap_ctci_diff_zct}). It is easy to
prove that $S D_{\rm max}^2 z/(1+S D_{\rm max} z)$ is monotonically
increasing with $z$, and $1/z$ is monotonically decreasing with $z$,
therefore
\begin{subequations}
\begin{equation} \label{cap_ctci_diff_res1}
\int_0^{z_{\rm t}} \frac{S D_{\rm max}^2 z} {1+S D_{\rm max} z}
f_{z_{\rm eff}}(z) dz< \int_0^{z_{\rm t}} \frac{S D_{\rm max}^2
z_{\rm t}} {1+S D_{\rm max} z_{\rm t}} f_{z_{\rm eff}}(z) dz=
\frac{S D_{\rm max}^2 z_{\rm t}} {1+S D_{\rm max} z_{\rm t}}
F_{z_{\rm eff}}(z_{\rm t})
\end{equation}
\begin{equation} \label{cap_ctci_diff_res2}
\int_{z_{\rm t}}^{\infty} \frac{1}{z} f_{z_{\rm eff}}(z) dz<
\int_{z_{\rm t}}^{\infty} \frac{1}{z_{\rm t}} f_{z_{\rm eff}}(z)
dz=\frac{1-F_{z_{\rm eff}}(z_{\rm t})}{z_{\rm t}}
\end{equation}
\end{subequations}
Multiplying (\ref{cap_ctci_diff_res1}) with
(\ref{cap_ctci_diff_res2}) then substituting the result into
(\ref{cap_ctci_diff_zct}), we obtain $d C_{\rm CTCI}/d z_{\rm t}
>0$, i.e. $C_{\rm CTCI}$ is monotonically increasing with $z_{\rm
t}$ regardless of $S$.

\section{Proof of Theorem \ref{slope1_thm}} \label{slope1_condt}

Let $C=\textrm{E}[\log (1+S D(z_{\rm eff}) z_{\rm eff})]=
\int_0^{\infty} \log (1+S D(z) z) f_{z_{\rm eff}}(z) dz$ and denote
the complementary set of $\mathscr{O}= \{ z| D(z)=0 \}$ on
$[0,\infty)$ by $\mathscr{\overline{O}}$. We have
\begin{equation} \label{prelog_raw}
\frac{dC}{d(\log S)}=S \frac{dC}{dS}= \int_0^{\infty} \left( \frac{S
D(z) z}{1+S D(z) z} \right) f_{z_{\rm eff}}(z) dz
\end{equation}
Clearly, $\lim_{S \rightarrow \infty} (S D(z) z)/(1+S D(z) z)$
becomes $1$ with $D(z) \ne 0$ and $0$ with $D(z)=0$, and the limit
and integral can be exchanged since the absolute value of the
integrand is upper bounded by $f_{z_{\rm eff}}(z)$, which is an
integrable function. Therefore the limit of (\ref{prelog_raw}) as $S
\rightarrow \infty$ becomes
\begin{equation}
\begin{split}
\lim_{S \rightarrow \infty} \frac{dC}{d(\log S)} &=
\int_{\mathscr{\overline{O}}} 1 \cdot f_{z_{\rm eff}}(z) dz+
\int_{\mathscr{O}} 0 \cdot f_{z_{\rm eff}}(z) dz= \int_0^{\infty}
f_{z_{\rm eff}}(z) dz- \int_{\mathscr{O}} f_{z_{\rm eff}}(z) dz+ 0\\
&= 1-\textrm{Pr}(\mathscr{O}).
\end{split}
\end{equation}

\section{Proof of Lemma \ref{lem_int}} \label{lem_int_proof}

With $\mu$ being a positive constant, $\log ((1+S\mu)/(S\mu))$ is
independent of $z$, therefore
\begin{equation}
\lim_{S \rightarrow \infty} \int_{0}^{\infty} \log \left(
\frac{1+S\mu}{S\mu} \right) f_{z_{\rm eff}}(z) dz= \lim_{S
\rightarrow \infty} \log \left( \frac{1+S\mu}{S\mu} \right) \cdot
\int_{0}^{\infty} f_{z_{\rm eff}}(z) dz= \log 1 \cdot 1= 0
\end{equation}
Since $\log ((1+Sz)/S)= \log (1/S+z)$ is monotonically decreasing
with $S$, it is easy to show that $\int_{0}^{\infty} \log ((1+Sz)/S)
f_{z_{\rm eff}}(z) dz$ is monotonically decreasing with $S$ and has
an infimum of $\int_{0}^{\infty} \log z f_{z_{\rm eff}}(z) dz$ as $S
\rightarrow \infty$. Consequently, due to monotone convergence
theorem, which indicates that the limit of a sequence of real
numbers is its infimum if it is decreasing and bounded below, we
have
\begin{equation}
\begin{split}
\lim_{S \rightarrow \infty} \int_{0}^{\infty} \log \left(
\frac{1+Sz}{S} \right) f_{z_{\rm eff}}(z) dz &= \inf \left\{
\int_{0}^{\infty} \log \left( \frac{1+Sz}{S} \right)
f_{z_{\rm eff}}(z) dz: S>0 \right\}\\
&= \int_{0}^{\infty} \log z f_{z_{\rm eff}}(z) dz
\end{split}
\end{equation}
and therefore
\begin{equation}
\lim_{S \rightarrow \infty} \int_{0}^{\infty} \log \left(
\frac{1+Sz}{Sz} \right) f_{z_{\rm eff}}(z) dz= \lim_{S \rightarrow
\infty} \int_{0}^{\infty} \log \left( \frac{1+Sz}{S} \right)
f_{z_{\rm eff}}(z) dz- \int_{0}^{\infty} \log z f_{z_{\rm eff}}(z)
dz=0
\end{equation}

\section{Proof of Theorem \ref{th_gaps}} \label{th_gaps_proof}

We first prove (\ref{gap_oa_ra}), which will be useful for
subsequent derivations since the gaps of OA and RA with respect to a
third scheme become equivalent. From (\ref{cap2_cpra}) and
(\ref{cap2_oa}) we obtain
\begin{equation} \label{diff_oa_cpra}
\begin{split}
C_{\rm OA}-C_{\rm RA} &\leq \int_{z_{\rm t}}^{\infty} \log \left(
\frac{z}{z_{\rm t}} \right) f_{z_{\rm eff}}(z) dz- \int_{z_{\rm
t}}^{\infty} \log (1+Sz) f_{z_{\rm eff}}(z)
dz\\
&= \int_{z_{\rm t}}^{\infty} \log \left( \frac{z}{z_{\rm t}+ Sz_{\rm
t}z} \right) f_{z_{\rm eff}}(z) dz< \int_{z_{\rm t}}^{\infty} \log
\left( \frac{z+S^{-1}}{z_{\rm t}+ Sz_{\rm
t}z} \right) f_{z_{\rm eff}}(z) dz\\
&=\int_{z_{\rm t}}^{\infty} \log \left( \frac{1}{Sz_{\rm t}} \right)
f_{z_{\rm eff}}(z) dz= -\log (Sz_{\rm t}) (1-F_{z_{\rm eff}} (z_{\rm
t}))
\end{split}
\end{equation}
Since the OA threshold satisfies $\lim_{S \rightarrow \infty} z_{\rm
t}=0$ by Lemma \ref{lem_zt}, $\lim_{S \rightarrow \infty} Sz_{\rm
t}=\lim_{S \rightarrow \infty} d C_{\rm OA}/d (\log S)=1$ based on
(\ref{oa_slope1}). Furthermore, $C_{\rm OA}-C_{\rm RA} \geq 0$ due
to the optimality of OA. Taking the limit in (\ref{diff_oa_cpra}),
we obtain (\ref{gap_oa_ra}). This indicates that RA is
asymptotically optimal with sufficiently high average SNR.

We now use Lemma \ref{lem_int} and (\ref{gap_oa_ra}) to show
(\ref{gap_awgn_oa})
\begin{equation} \label{gap2_awgn_oa}
\begin{split}
\lim_{S \rightarrow \infty} (C_{\rm AWGN}-C_{\rm OA}) &= \lim_{S
\rightarrow \infty} (C_{\rm AWGN}-C_{\rm RA})= \lim_{S \rightarrow
\infty} \int_0^{\infty} \log \left(
\frac{1+S\textrm{E}[z_{\rm eff}]} {1+Sz} \right) f_{z_{\rm eff}}(z) dz\\
&=^{\dagger} \lim_{S \rightarrow \infty} \int_0^{\infty} \log \left(
\frac{1+S\textrm{E}[z_{\rm eff}]} {1+Sz} \right) f_{z_{\rm eff}}(z)
dz+ \lim_{S \rightarrow \infty} \int_0^{\infty} \log \left(
\frac{1+Sz} {Sz} \right) f_{z_{\rm eff}}(z) dz\\
& +\lim_{S \rightarrow \infty} \int_0^{\infty} \log \left(
\frac{S\textrm{E}[z_{\rm eff}]} {1+S\textrm{E}[z_{\rm
eff}]} \right) f_{z_{\rm eff}}(z) dz\\
&= \int_0^{\infty} \log \left( \frac{\textrm{E}[z_{\rm eff}]}{z}
\right) f_{z_{\rm eff}}(z) dz= \log (\textrm{E}[z_{\rm eff}])-
\textrm{E}[\log z_{\rm eff}]
\end{split}
\end{equation}
where we use (\ref{lem_int_eqn2}) in the second term, and
(\ref{lem_int_eqn1}) with $\mu=\textrm{E}[z_{\rm eff}]$ in the third
term following the third equality (marked ``$\dagger$''). Notice
that (\ref{gap_awgn_oa}) is guaranteed to be finite when {\bf A1}
holds, since it ensures $C_{\rm AWGN}$ and $C_{\rm OA}$ have the
same high-SNR pre-log constant, and also a finite $\textrm{E}[\log
z_{\rm eff}]$. Moreover, we have $\log (\textrm{E}[z_{\rm eff}])
\geq \textrm{E}[\log z_{\rm eff}]$ based on Jensen's inequality and
thus (\ref{gap_awgn_oa}) is non-negative.

To show (\ref{gap_oa_ci}), using a similar approach as in
(\ref{gap2_awgn_oa}), we obtain
\begin{equation} \label{gap2_oa_ci}
\begin{split}
\lim_{S \rightarrow \infty} (C_{\rm OA}-C_{\rm CI}) &=^{\dagger}
\lim_{S \rightarrow \infty} (C_{\rm RA}-C_{\rm CI})= \lim_{S
\rightarrow \infty} \left( \int_0^{\infty} \log (1+Sz) f_{z_{\rm
eff}}(z)
dz- \log (1+S\textrm{E}^{-1} [z_{\rm eff}^{-1}]) \right)\\
&= \lim_{S \rightarrow \infty} \int_0^{\infty} \log \left(
\frac{1+Sz} {1+S\textrm{E}^{-1} [z_{\rm eff}^{-1}]} \right)
f_{z_{\rm eff}}(z) dz=^{\ddagger} \int_0^{\infty} \log \left(
\frac{z} {\textrm{E}^{-1} [z_{\rm eff}^{-1}]} \right) f_{z_{\rm eff}}(z) dz\\
&= \textrm{E}[\log z_{\rm eff}]+ \log (\textrm{E}[z_{\rm eff}^{-1}])
\end{split}
\end{equation}
where the first equality (marked ``$\dagger$'') is based on
(\ref{gap_oa_ra}), and the fourth equality (marked ``$\ddagger$'')
is based on Lemma \ref{lem_int} with $\mu=\textrm{E}^{-1} [z_{\rm
eff}^{-1}]$. Similar to (\ref{gap_awgn_oa}), (\ref{gap_oa_ci}) is
non-negative due to Jensen's inequality, and it is finite since both
$\textrm{E}[\log z_{\rm eff}]$ and $\log (\textrm{E}[z_{\rm
eff}^{-1}])$ are finite given {\bf A2}.

\section{Proof of Theorem \ref{th_lowsnr}} \label{th_lowsnr_proof}

Since $C_{\rm CI} \leq C_{\rm CTCI} \leq C_{\rm RA}$ for any $S$,
and all capacities are zero at $S=0$, the inequalities in
(\ref{lowsnr_comp1}) hold, and the first and last equalities follow
from (\ref{slope_cpra}) and (\ref{slope_ci}). (\ref{lowsnr_comp3})
can be obtained based on the derivation after Lemma \ref{lem_zt} in
\ref{asympt_slopes}, and the fact that the least upper bound is no
less than the mean. (\ref{lowsnr_comp2}) holds since
(\ref{cap_otci_diff_p0}) indicates that $\lim_{S \rightarrow 0}
dC_{\rm TCI}/dS$ is monotonically increasing with $z_{\rm t}$. In
addition, $\lim_{S \rightarrow 0} dC_{\rm TCI}/dS \geq z_{\rm t}$ as
can be obtained through upper-bounding the denominator in
(\ref{cap_otci_diff_p0}) by $\int_{z_{\rm t}}^{\infty} z_{\rm
t}^{-1} f_{z_{\rm eff}}(z) dz$, therefore $z_{\rm t} \geq \textrm{E}
[z_{\rm eff}]$ is a sufficient condition for $\lim_{S \rightarrow 0}
dC_{\rm TCI}/dS \geq \lim_{S \rightarrow 0} dC_{\rm AWGN}/dS$.

\bibliographystyle{IEEEtran}
\bibliography{reflist}

\begin{thebibliography}{10}
\providecommand{\url}[1]{#1}
\csname url@rmstyle\endcsname
\providecommand{\newblock}{\relax}
\providecommand{\bibinfo}[2]{#2}
\providecommand\BIBentrySTDinterwordspacing{\spaceskip=0pt\relax}
\providecommand\BIBentryALTinterwordstretchfactor{4}
\providecommand\BIBentryALTinterwordspacing{\spaceskip=\fontdimen2\font plus
\BIBentryALTinterwordstretchfactor\fontdimen3\font minus
  \fontdimen4\font\relax}
\providecommand\BIBforeignlanguage[2]{{%
\expandafter\ifx\csname l@#1\endcsname\relax
\typeout{** WARNING: IEEEtran.bst: No hyphenation pattern has been}%
\typeout{** loaded for the language `#1'. Using the pattern for}%
\typeout{** the default language instead.}%
\else
\language=\csname l@#1\endcsname
\fi
#2}}

\bibitem{goldsmith97}
A.~J. Goldsmith and P.~P. Varaiya, ``{Capacity of fading channels with channel
  side information},'' \emph{IEEE Transactions on Information Theory}, vol.~43,
  no.~6, pp. 1986--1992, Nov. 1997.

\bibitem{laourine08}
A.~Laourine, M.-S. Alouini, S.~Affes, and A.~Stephenne, ``{On the capacity of
  generalized-k fading channels},'' \emph{IEEE Transactions on Wireless
  Communications}, vol.~7, no.~7, pp. 2441--2445, Jul. 2008.

\bibitem{laourine09}
------, ``{On the performance analysis of composite multipath/shadowing
  channels using the G-distribution},'' \emph{IEEE Transactions on
  Communications}, vol.~57, no.~4, pp. 1162--1170, Apr. 2009.

\bibitem{song05}
G.~Song and Y.~Li, ``{Asymptotic throughput analysis of multiuser diversity},''
  \emph{IEEE Global Telecommunications Conference}, vol.~3, pp. 1289--1293,
  Nov. 2005.

\bibitem{alouini97}
M.-S. Alouini and A.~J. Goldsmith, ``{Capacity of Nakagami multipath fading
  channels},'' \emph{IEEE 47th Vehicular Technology Conference}, vol.~1, pp.
  358--362, May. 1997.

\bibitem{alouini99}
------, ``{Capacity of Rayleigh fading channels under different adaptive
  transmission and diversity-combining techniques},'' \emph{IEEE Transactions
  on Vehicular Technology}, vol.~48, no.~4, pp. 1165--1181, Jul. 1999.

\bibitem{shao99}
J.~W. Shao, M.-S. Alouini, and A.~J. Goldsmith, ``{Impact of fading correlation
  and unequal branch gains on the capacity of diversity systems},'' \emph{IEEE
  49th Vehicular Technology Conference}, vol.~3, pp. 2159--2163, May. 1999.

\bibitem{bithas09}
P.~S. Bithas and P.~T. Mathiopoulos, ``{Capacity of Correlated Generalized
  Gamma Fading With Dual-Branch Selection Diversity},'' \emph{IEEE Transactions
  on Vehicular Technology}, vol.~58, no.~9, pp. 5258--5663, Nov. 2009.

\bibitem{mallik04}
R.~K. Mallik, M.~Z. Win, J.~W. Shao, M.-S. Alouini, and A.~J. Goldsmith,
  ``{Channel capacity of adaptive transmission with maximal ratio combining in
  correlated Rayleigh fading},'' \emph{IEEE Transactions on Wireless
  Communications}, vol.~3, no.~4, pp. 1124--1133, Jul. 2004.

\bibitem{khatalin07}
S.~Khatalin and J.~P. Fonseka, ``{Channel capacity of dual-branch diversity
  systems over correlated Nakagami-m fading with channel inversion and fixed
  rate transmission scheme},'' \emph{IET Communications}, vol.~1, pp.
  1161--1169, Dec. 2007.

\bibitem{hussain09}
Z.~M. Hussain, S.~A. Zummo, F.~S. Al-Qahtani, and A.~K. Gurung, ``{Spectral
  efficiency evaluation for selection combining diversity (SCD) scheme over
  slow fading},'' \emph{IET Communications}, vol.~3, pp. 1443--1451, Sep. 2009.

\bibitem{goldsmith97b}
A.~J. Goldsmith and S.-G. Chua, ``{Variable-rate variable-power MQAM for fading
  channels},'' \emph{IEEE Transactions on Communications}, vol.~45, no.~10, pp.
  1218--1230, Oct. 1997.

\bibitem{vishwanath03}
S.~Vishwanath and A.~J. Goldsmith, ``{Adaptive turbo-coded modulation for
  flat-fading channels},'' \emph{IEEE Transactions on Communications}, vol.~51,
  no.~6, pp. 964--972, Jun. 2003.

\bibitem{fellerbook71}
W.~Feller, \emph{{An Introduction to Probability Theory and Its Applications:
  Volume II}}, 2nd~ed.\hskip 1em plus 0.5em minus 0.4em\relax New York: John
  Wiley and Sons, 1971.

\bibitem{wang03}
Z.~Wang and G.~B. Giannakis, ``{A simple and general parameterization
  quantifying performance in fading channels},'' \emph{IEEE Transactions on
  Communications}, vol.~51, no.~8, pp. 1389--1398, Aug. 2003.

\bibitem{mceliece84}
R.~McEliece and W.~Stark, ``{Channels with block interference},'' \emph{IEEE
  Transactions on Information Theory}, vol.~30, no.~1, pp. 44--53, Jan. 1984.

\bibitem{lapidoth05}
A.~Lapidoth, ``{On the asymptotic capacity of stationary Gaussian fading
  channels},'' \emph{IEEE Transactions on Information Theory}, vol.~51, no.~2,
  pp. 437--446, Feb. 2005.

\bibitem{alouini00}
M.-S. Alouini and A.~J. Goldsmith, ``{Comparison of fading channel capacity
  under different CSI assumptions},'' \emph{IEEE 52nd Vehicular Technology
  Conference}, vol.~4, pp. 1844--1849, Sep 2000.

\bibitem{tsebook05}
D.~N.~C. Tse and P.~Viswanath, \emph{{Fundamentals of Wireless Communication}},
  1st~ed.\hskip 1em plus 0.5em minus 0.4em\relax Cambridge: Cambridge
  University Press, Jun. 2005.

\bibitem{biglieri98}
E.~Biglieri, J.~Proakis, and S.~Shamai, ``{Fading Channels:
  Information-Theoretic and Communications Aspects},'' \emph{IEEE Transactions
  on Information Theory}, vol.~44, no.~6, pp. 2619--2692, Oct. 1998.

\bibitem{leong10}
A.~S. Leong and S.~Dey, ``{On Scaling Laws of Diversity Schemes in
  Decentralized Estimation},'' {accepted for publication on \emph{IEEE
  Transactions on Information Theory}, also available online at
  http://arxiv.org/abs/1002.4473}.

\bibitem{dasguptabook08}
A.~DasGupta, \emph{{Asymptotic Theory of Statistics and Probability}},
  1st~ed.\hskip 1em plus 0.5em minus 0.4em\relax Springer, Mar. 2008.

\bibitem{pun07}
M.-O. Pun, V.~Koivunen, and H.~V. Poor, ``{Opportunistic Scheduling and
  Beamforming for MIMO-SDMA Downlink Systems with Linear Combining},''
  \emph{IEEE 18th International Symposium on Personal, Indoor and Mobile Radio
  Communications}, pp. 1--6, Sep. 2007.

\end{thebibliography}

\end{document}